\documentclass[conference]{IEEEtran}

\usepackage[inline]{enumitem}
\usepackage{hyperref}
\usepackage{cleveref}
\usepackage{url}
\usepackage{xcolor}
\usepackage{listings}
\PassOptionsToPackage{table}{xcolor}
\PassOptionsToPackage{dvipsnames}{xcolor}
\usepackage{colortbl}
\usepackage[markup=underlined,final]{changes}
\usepackage{comment}
\usepackage{amssymb}
\usepackage{amsthm}
\usepackage{booktabs}
\usepackage{tabularx}
\usepackage{multirow}
\usepackage{longtable}
\usepackage{microtype}
\usepackage{pifont}

\usepackage{tikz}
\newcommand*\emptycirc[1][1ex]{\tikz\draw[thick] (0,0) circle (#1);}
\newcommand*\halfcirc[1][1ex]{
  \begin{tikzpicture}
  \draw[fill] (0,0)-- (90:#1) arc (90:270:#1) -- cycle ;
  \draw[thick] (0,0) circle (#1);
  \end{tikzpicture}}
\newcommand*\fullcirc[1][1ex]{\tikz\fill (0,0) circle (#1);}

\definecolor{ltblue}{rgb}{0,0.4,0.4}
\definecolor{dkblue}{rgb}{0,0.1,0.6}
\definecolor{dkgreen}{rgb}{0,0.35,0}
\definecolor{dkviolet}{rgb}{0.3,0,0.5}
\definecolor{dkred}{rgb}{0.5,0,0}

\usepackage{rotating}
\usepackage{lstcoq}
\usepackage{lipsum}
\usepackage{todonotes}

\setlist[itemize]{itemsep=3pt, parsep=0pt, partopsep=5pt, topsep=3pt}
\setlist[enumerate]{itemsep=3pt, parsep=0pt, partopsep=5pt, topsep=3pt}

\newcommand{\ie}{i.e.}
\newcommand{\eg}{e.g.}
\newcommand{\etal}{et al.}

\newcommand{\cyes}{\fullcirc}
\newcommand{\cno}{\emptycirc}

\newcommand{\alert}[1]{{\color{red} #1 }}

\newcommand*\important[1]{\tikz[baseline=(char.base)]{
    \node[shape=rectangle,fill=yellow!45,text=red,font=\bfseries,inner sep=0.4pt] (char) {\texttt{\underline{#1}}};}}

\definechangesauthor[color=violet]{LV}
\definechangesauthor[color=orange]{BF}
\definechangesauthor[color=teal]{PB}
\setcommentmarkup{\todo[color={authorcolor!20},size=\tiny]{#3: #1}}

\renewcommand{\alert}[1]{}
\renewcommand{\important}[1]{}

\newtheorem{innercustomthm}{Invariant}
\newenvironment{invariant}[1]
  {\renewcommand\theinnercustomthm{#1}\innercustomthm}
  {\endinnercustomthm}

\newcommand{\thetool}{our toolchain}
\newcommand{\themodel}{our model}
\newcommand{\framework}{WebSpec}
\newcommand{\theframework}{\framework}

\newcommand{\muz}{$\mu$Z}
\newcommand{\spacer}{Spacer}
\newcommand{\bmc}{BMC}

\newcommand{\code}[1]{\texttt{\hyphenchar\font=`\-\small#1}}
\renewcommand{\coqe}{\lstinline[language=Coq, breakatwhitespace, basicstyle=\small]}

\renewcommand{\autoref}{\Cref}

\newcommand{\bfparagraph}[1]{\smallskip\noindent\textbf{#1}}

\begin{document}
\title{\framework{}: Towards Machine-Checked Analysis of Browser Security Mechanisms}

\author{
  \IEEEauthorblockN{Lorenzo Veronese, Benjamin Farinier, Pedro Bernardo, Mauro Tempesta, Marco Squarcina, Matteo Maffei}
  \IEEEauthorblockA{TU Wien}
} 

\maketitle

\begin{abstract}

The complexity of browsers has steadily increased over the years, driven by the continuous introduction and update of Web platform components, such as novel Web APIs and security mechanisms. Their specifications are manually reviewed by experts to identify potential security issues. However, this process has proved to be error-prone due to the extensiveness of modern browser specifications and the interplay between new and existing Web platform components.
To tackle this problem, we developed \framework{}, the first formal security framework for the analysis of browser security mechanisms, which enables both the automatic discovery of logical flaws and the development of machine-checked security proofs. \framework, in particular, includes a comprehensive semantic model of the browser in the Coq proof assistant, a formalization in this model of ten Web security invariants, and a toolchain turning the Coq model and the Web invariants into SMT-lib formulas to enable model checking with the Z3 theorem prover. If a violation is found, the toolchain automatically generates executable tests corresponding to the discovered attack trace, which is validated across major browsers.

We showcase the effectiveness of \framework{} by discovering two new logical flaws caused by the interaction of different browser mechanisms and by identifying three previously discovered logical flaws in the current Web platform, as well as five in old versions. Finally, we show how \framework{} can aid the verification of our proposed changes to amend the reported inconsistencies affecting the current Web platform.
\end{abstract}

\section{Introduction}
Web browsers are considered among the most complex software in use today, and the number of Web platform components, i.e., browser functionalities and security mechanisms, is constantly increasing. These are typically proposed by browser vendors in the form of a W3C \emph{Editor's Draft} and discussed within the community. If enough consensus is reached, the standardization process has to progress through several \emph{maturity levels} before becoming a W3C recommendation. 

While the implementation of new Web platform components is subject to extensive compliance testing (see, e.g., the \emph{Web Platform Tests} project~\cite{WPT}), their specifications undergo a  manual expert review to identify potential issues: this is a continuous and extremely complex process that has to consider the interplay with legacy APIs and should, in principle, be revised whenever new components land on the Web platform.

Unfortunately, manual reviews tend to overlook logical flaws, eventually leading to critical security vulnerabilities.
For example, the \texttt{HttpOnly} flag was introduced by Internet Explorer 6~\cite{OWASPCookies} as a way to protect the confidentiality of cookies with this attribute by not exposing them to scripts.
Eight years after its launch, Singh~\etal{} discovered that this property could be trivially violated by any scripts accessing the response headers of an AJAX request via the \texttt{getResponseHeader} function~\cite{SinghMWL10}. Security vulnerabilities at the level of Web specifications have also affected CORS~\cite{AkhaweBLMS10}, CSP~\cite{SomeBR17}, and Trusted Types~\cite{TTSecureCtx}, to name a few. 

We argue that this dire situation stems from several concurring factors:
\begin{enumerate*}[label=(\roman*)]
  \item Web platform components are specified informally and therefore their analysis, albeit conducted by expert eyes, may easily overlook corner cases;
  \item there is no precise understanding of which security properties should be seen as invariants in the Web and, thus, be preserved by updates of the Web platform;
  \item Web platform components are typically evaluated in isolation, without considering their interactions, that is, the entangled nature of the Web platform.
\end{enumerate*}

\begin{figure}[t]
  \centering
  \includegraphics[width=0.35\textwidth]{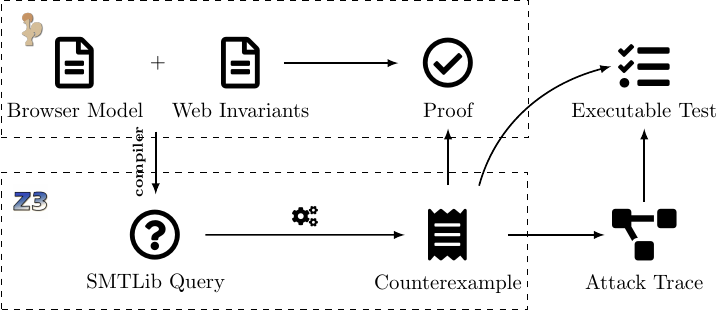}
  \caption{The \framework{} framework}
  \label{fig:toolchain}
\end{figure}
\bfparagraph{Our Contributions.}
In this work, we advocate a paradigm shift, letting Web platform components and their interplay undergo a formal security analysis as opposed to a manual expert review.
In particular, we introduce \framework{}, the first formal framework for the security analysis of browser security mechanisms that supports the automated detection of logical flaws as well as machine-checked security proofs. As outlined in \autoref{fig:toolchain}, \framework{} includes:

\begin{itemize}
\item a formal browser model in Coq (\autoref{sec:model}), which (i) formalizes a core set of Web platform components, including well-established  (cookies, SOP, CORS, etc.) as well as recent ones (e.g., CSP level 3 and Trusted Types), and (ii) supports the definition of Web invariants, i.e., properties that are expected to hold in the Web;
\item the \framework{} verification toolchain (\autoref{sec:compiler}), encompassing a compiler and a trace verifier.  
The compiler translates the browser model and the Web invariants into SMT-lib formulas to enable model checking by the Z3 automated theorem prover. A salient feature of \framework{} is the support for both bug finding and proof generation.  If a violation is found, Z3 reconstructs the minimal sequence of actions leading to it, and the trace verifier displays the corresponding attack trace and maps it to executable tests in order to systematically validate the Web inconsistencies found in the model on major browsers. For Web invariants that instead hold,  proofs can be directly derived by Z3 against the SMT-lib encoding or   manually written and machine-checked in Coq. 
\end{itemize}
We demonstrate the effectiveness of \framework{} by:
\begin{itemize}
\item defining ten Web invariants against which we  identify (i) a new attack on cookies caused by the interaction with legacy APIs,  (ii) a new inconsistency between CSP and a planned change to the HTML standard, as well as  (iii) three previously reported logical flaws in the current Web platform (\autoref{sec:webinvariants});
\item validating all five Web inconsistencies against the latest versions of Chrome and Firefox;
\item adjusting the model to reflect past states of the Web platform in order to  identify five previously published attacks, with the goal of showing that automated security analysis would have prevented these vulnerabilities; 
\item writing  the proofs in the Coq model for the correctness of the  four fixes we propose against the  vulnerabilities on the current Web platform (\autoref{sec:proof});
\item conducting an experimental evaluation to demonstrate the effectiveness of the \framework{} toolchain and the optimizations we integrated therein (\autoref{sec:evaluation}); 
\item systematically analyzing the state-of-the-art in formal browser models, showing that the model presented in this work is the most comprehensive one in terms of supported client-side security mechanisms (\autoref{sec:related}). 
\end{itemize}


\section{Browser Model}
\label{sec:model}
This section provides an overview of the main components of our browser model written in Coq. The model focuses on Web platform components, i.e., browser functionalities and security mechanisms,  abstracting away from the network and Web servers.
Our formalization enables reasoning about all possible sequences of events leading to an inconsistent state without necessarily having to model a specific Web application.
We are indeed interested in proving and disproving Web invariants, i.e., \emph{properties of the Web platform that are expected to hold across its updates and independently on how its components can interact with each other}~\cite{AkhaweBLMS10}. Web invariants are supposed to hold for all Web applications, irrespectively of application-specific assumptions that attackers could violate. For instance, scripts in our model can, in principle, execute arbitrary sequences of any of the API calls we support, as this would be the case in presence of cross-site scripting attacks. The model also includes configuration flags that enable reasoning on former states of the Web platform or testing new proposals prior to their implementation.

\subsection{Core Abstractions}
\label{sec_global_event_state}

\begin{figure*}[t]
  \centering
  \includegraphics[width=\linewidth]{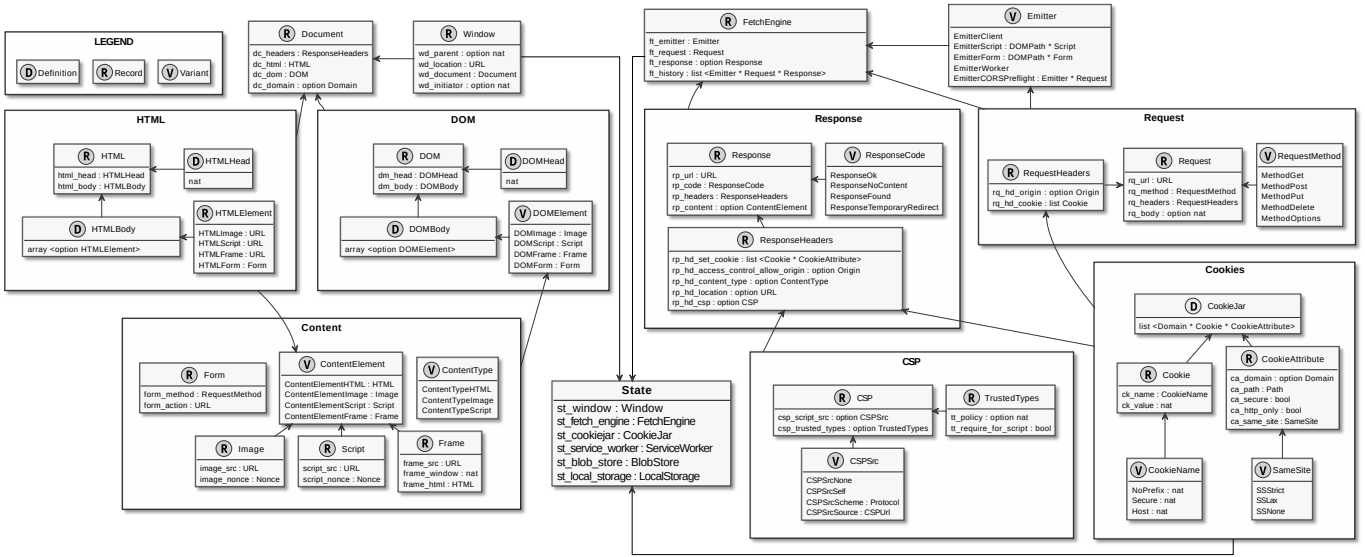}
  \caption{Browser \texttt{State} and overview of its components ($\to$: referenced by): rendering components are on the left of the state, networking on the right}
  \label{fig:model}
\end{figure*}

The browser is modeled as a transition system in which a state evolves from an initial to a
final configuration following a list of events and according to an inductive
relation named \coqe{Reachable} parameterized by a global environment:
\begin{lstlisting}[language=coq, numbers=none]
Inductive Reachable : Global -> list Event -> State -> Prop.
\end{lstlisting}
Intuitively, given a global environment \coqe{gb}, a list of events \coqe{evs}, and a state \coqe{st}, \coqe{Reachable gb evs st} means that, starting from a given initial state, \coqe{st} is reachable by executing sequentially the events in \coqe{evs} under environment \coqe{gb}.

The \coqe{Global} environment contains concrete values (\eg,\ the browser configuration) or symbolic variables (\eg,\ a set of pages) which are constant through the evolution of the browser state.
An \coqe{Event} represents an atomic action that modifies the state, e.g., sending a network request or updating the DOM, which may originate from different sources, such as the browser itself, a script, or a service worker.
A \coqe{State} (\autoref{fig:model}) is a collection of datatypes used to model browser components. 

Based on these ingredients, we formalize Web invariants within our model as follows, where \coqe{hypothesis} and \coqe{conclusion} are predicates that may refer to the global environment, past events, or the current state of the browser:
\begin{lstlisting}[language=coq]
Parameter hypothesis : Global -> list Event -> State -> Prop.
Parameter conclusion : Global -> list Event -> State -> Prop.

Definition Invariant (gb: Global) (evs: list Event) (st: State) := Reachable gb evs st -> hypothesis gb evs st -> conclusion gb evs st.
\end{lstlisting}
In \autoref{sec:webinvariants}, we introduce ten Web invariants and discuss their security implications caused by the interplay of different Web platform components.

\subsection{Page Rendering}
\label{sec_window_document_html_dom}

The main component used to model the rendering functionality is the
\coqe{Window} datatype.
 \coqe{Window} represents a window in terms of browsing context~\cite[§7.1]{HTML},
\ie, an environment in which the browser displays a document.
The field \coqe{wd_location} is the URL being visited and \coqe{wd_document}
contains the displayed document.
Since a \coqe{Window} can represent either a top-level window or a frame,
\coqe{wd_parent} contains an optional index which, if empty, denotes a
top-level window or points to the parent frame otherwise.
Similarly, \coqe{wd_initiator} contains an optional index which is used to
track the source browsing context of this window~\cite[§7.11]{HTML} by storing a reference to the window responsible for starting the navigation.

The \coqe{Document} datatype represents a Web page loaded and rendered in a browser window.
When a page is loaded, \coqe{dc_html} represents the HTML code of the response, while \coqe{dc_dom} contains the rendered elements of the page. Static elements, \eg, forms and possibly other markup tags, are rendered immediately. Subresources of the page, such as frames and scripts, require an additional request to be included in \coqe{dc_dom}.
For instance, the presence of a \coqe{HTMLFrame} in \coqe{dc_html} might cause three additional events to be executed in sequence: a request (\coqe{EvRequest}), a response (\coqe{EvResponse}), followed by the update of the DOM (\coqe{EvDOMUpdate}) resulting in \coqe{DOMFrame} being added to \coqe{dc_dom}.
This approach enables fine-grained modeling of the rendering process of the browser. In particular, our model  captures the order in which resources are loaded and the presence or absence of specific elements.

\framework{} currently supports forms with the \code{method} and \code{action} attributes, images, scripts, and frames. Rendered frames in \coqe{dc_dom} contain a reference to the corresponding \coqe{Window} (\autoref{fig:model}), reflecting the tree-like structure of the DOM.
Finally, \coqe{Document} also includes the list of headers (\coqe{dc_headers}) in the HTTP response used to render the page and \coqe{dc_domain}, an optional field used to model domain relaxation via the \code{document.domain} API (\autoref{sec_javascript}).

\subsection{Networking and Cookies}
\label{sec_fetchengine_request_response_cookie}

The main component used to model the networking functionality is the
\coqe{FetchEngine}, which abstracts network access and is responsible for sending
requests and receiving responses.
\coqe{ft_request} contains the last emitted request,
and \coqe{ft_emitter} maps to the originator of the request, \ie, whether the request is top-level or generated by the inclusion of subresources, issued by a script, a form, a worker, or it is a CORS preflight.
\coqe{ft_response} is a field that either contains the corresponding response or is empty if the request is still pending.
Finally, we store \coqe{Emitter * Request * Response} triples in
\coqe{ft_history} in order to keep track of previous network accesses.

The modeling of \coqe{Request} and \coqe{Response} is rather straightforward,
as shown in \autoref{fig:model}.
We support requests and responses through HTTP and HTTPS protocols.
For requests, we model the HTTP methods \code{GET}, \code{POST}, \code{PUT},
\code{DELETE}, and \code{OPTION}.
Concerning responses, the following HTTP status codes are supported:
\begin{enumerate*}[label=(\roman*)]
  \item \code{200 OK}, successful response,
  \item \code{204 No Content}, successful response with an empty body,
  \item \code{302 Found}, redirection with no integrity guarantees in the redirected request over the HTTP method and the body of the original request~\cite{MDN302},
  \item \code{307 Temporary Redirect}, redirection enforcing that the method and body of the original request are preserved in the redirected one.
\end{enumerate*}

Supported headers are \code{Origin}, \code{Cookie} and \code{Referer} for requests, and \code{Content-Type}, \code{Set-Cookie}, \code{Location} and \code{Referrer-Policy} for responses. To support CSP and CORS, we include the \code{Content-Security-Policy} and \code{Access-Control-Allow-Origin} headers.

Cookies are stored in the \coqe{CookieJar} as a list of triples in the form \coqe{Domain * Cookie * CookieAttribute}, where \coqe{Domain} represents the host setting the cookie, \coqe{Cookie} is a pair corresponding to the name of the cookie and its value, and \coqe{CookieAttribute} is a record containing the attributes. Modeled cookie attributes are \code{Domain}, \code{Path}, \code{Secure}, \code{HttpOnly}, and \code{SameSite}. We also support cookie prefixes which enforce additional constraints~\cite{RFC6265bis}:
\begin{enumerate*}[label=(\roman*)]
\item \texttt{\_\_Secure-} cookies must be set with the \texttt{Secure} attribute and from a page served over HTTPS;
\item \texttt{\_\_Host-} cookies have all the constraints of the \texttt{\_\_Secure-} attribute, plus the \texttt{Path} attribute must be set to the value ``\texttt{/}'' (ensuring that cookies will be attached to all requests) and must not contain a \texttt{Domain} attribute, thus restricting the scope to the host that set it.
\end{enumerate*}

\subsection{Additional Features}
\label{sec_additional_features}

Starting from the core functionalities discussed in the previous sections, our model can be extended to support other Web components, including novel security mechanisms that could benefit from the automated formal analysis enabled by \framework{}. We discuss in the following six additional Web components supported by our model.

\begin{description}[style=unboxed,leftmargin=0cm]
  \item[Content Security Policy.]
  The \emph{Content Security Policy} (CSP) allows Web developers to tighten the security of Web applications by controlling which resources can be loaded and executed by the browser.
  Originally, the CSP was designed to mitigate content injection vulnerabilities. Subsequently, it was extended to restrict browser navigation (\eg, \texttt{form-action}, \texttt{frame-ancestors}) and protect DOM XSS sinks (via \texttt{trusted-types}).
  A CSP policy consists of a set of directives and source expressions specifying an allow-list of actions the page is allowed to perform.
  Currently modeled CSP directives are: \code{script-src}, which defines the allow-list for JavaScript sources, and \code{trusted-types} together with \code{require-trusted-types-for} for Trusted Types~\cite{TrustedTypesDraft} support, as explained later in this section.

  \item[Service Workers and Cache API.]
  A service worker~\cite{ServiceWorkers} acts as client-side proxy between Web applications and the network. Web applications are required to register a service worker, binding it to a specific origin and a scope. When enabled, a service worker can intercept and modify HTTP requests and corresponding responses initiated by its origin. Furthermore, service workers are activated also by cross-origin requests towards their registering origin. Using the Cache API, a service worker can store HTTP responses and serve them even when the network is unreachable. We reflect these capabilities in our model by considering a specific kind of service worker that can perform fetch requests, serve synthetic responses, and cache pairs of requests and responses, regardless of the scope. To this end, we also model a lightweight Cache API and assume that service workers have arbitrary access to it.
  \item[Local Scheme URLs.]
  We model requests to local scheme URLs~\cite{Fetch}, \ie, URLs with a local scheme such as \code{data:} and \code{blob:}, as virtual requests that do not generate a response from the network. We partially support the File API~\cite{FileAPI} by enabling the creation of blob URLs via the \code{URL.createObjectURL} JavaScript method. We assume that local URLs are accepted interchangeably with remote URLs, meaning that they can be navigated by frames, or included as a script in a page. We thoroughly discuss the interaction between Content Security Policy and local scheme URLs in \autoref{sec:csplocalschemes}.

  \item[Local Storage.]
  The Web Storage API~\cite[§12]{HTML} enables JavaScript to store and retrieve key/value pairs in the browser. The API provides two mechanisms to store data: \code{sessionStorage},  an ephemeral  storage that expires when the browser or the page is closed, and \code{localStorage}, which persists in the browser unless cleared explicitly. As we are interested in capturing single browser sessions, the difference between the two mechanisms is irrelevant. For this reason, we model only \code{localStorage}, providing methods to read and write data in the browser storage from any scripts.
  \item[Web Messaging.]
    Cross-origin communication is enabled by the postMessage API~\cite{HTML}. As we are interested in modeling messages that are sent and received---while we ignore messages that do not reach the destination---we encode the sending and receiving of a message as a single action. We also model the origin validation process performed in the receiving script. This way, we can capture potential security issues due to cross-origin messages processed without validating the sender's origin~\cite{SteffensS20}.

  \item[Trusted Types.]
  Trusted Types are an experimental security mechanism designed to prevent DOM XSS by restricting injection sinks to accepting only non-spoofable typed values in place of strings~\cite{TrustedTypesDraft}. These types can be created based on application-defined policies, allowing developers to specify via JavaScript a set of rules to protect injection sinks.
  Trusted Types are controlled by two CSP directives:
  \texttt{require-trusted-types-for 'script'}, which enables the enforcement of Trusted Types, \ie, instructs the browser to only accept Trusted Types for all DOM XSS injection sinks, and \texttt{trusted-types}, optionally followed by the name of one or more policies, which specifies the policies that are allowed to create Trusted Types objects.
  When no name is specified or when the special value \texttt{'none'} is used, no policy, and thus no Trusted Types, can be created, effectively disabling all DOM XSS sinks.
  We model the enforcement of Trusted Types on a page by mandating scripts to invoke the Trusted Type API to create a \code{TrustedHTML} object, and use it to modify the DOM via the \code{Element.innerHTML} property. Although we do not model the content of policies, we encode the ability to disallow the creation of any Trusted Types via the CSP directive \code{trusted-types 'none'}.
\end{description}

\subsection{JavaScript}
\label{sec_javascript}
Contrary to previous works~\cite{BohannonPhd}, instead of modeling scripts with an internal state and precise small-step semantics, we model them in terms of actions that the browser can perform. Since we are not interested in application-specific behavior, our abstraction captures the execution of sequences of Web API calls and the evolution of the browser's state.
For example, we model the fact that a script can set a cookie, or add a request-response pair to a cache using the Cache API, but we do not model how cookie data, requests, or responses are built. Instead, we introduce symbolic variables with constraints following the API specification.

Scripts in our model can update the DOM, set and get cookies using \code{Document.cookie}, or navigate frames using the \code{Window.location} setter.
They can use the Fetch API to perform network requests and read the corresponding responses up to SOP constraints, including support for CORS. We also model the legacy \code{Document.domain} API, which allows for cross-origin communication between windows in the same site by relaxing their \code{document.domain} property to a common ancestor. Although this API has been deprecated due to security concerns~\cite{MDNdocumentdomain}, and Google announced that it will be disabled by default starting from Chrome 109~\cite{ChromeDisableRelaxation}, it is still supported by all major browsers.
Scripts can also use the APIs described in \autoref{sec_additional_features}: they can update a cache from page context using \code{Cache.put}, communicate with other windows using \code{Window.postMessage}, create blob URLs with \code{URL.createObjectURL}, and create Trusted Types.

\section{\framework\ Toolchain}
\label{sec:compiler}

In the following, we present the \framework\ toolchain (\autoref{fig:toolchain}).
This toolchain comprises (i)
a compiler (\autoref{sec_toolchain_compiler}) which translates the browser
model and the Web invariants written in Coq into a query that can be
automatically checked by an SMT solver -- Z3 in the current implementation
(\autoref{sec_toolchain_solver})%
and (ii) a verifier which reconstructs from a SMT solution an attack trace enjoying
correctness and minimality, and validates it against real Web browser
(\autoref{sec_toolchain_verifier}).
Finally we discuss the advantages of our approach as  compared to prior work
(\autoref{sec_toolchain_discussion}).

\subsection{Compilation}
\label{sec_toolchain_compiler}

To automatically verify the (in)validity of our invariants, we
developed a compiler that translates the Coq model and the invariants into
SMT-lib formulas, which are then fed to the Z3 solver.
Technically, we compile Coq inductive types into CHC logic, \ie, first-order
logic with fixed-points expressed in terms of Constrained Horn
Clauses~\cite{HBM11,HB12}, in order to find inhabitants of the translated
inductive types, i.e., terms of these types.
In particular, the compiler translates  \coqe{Reachable} and all the inductive
types of kind \coqe{Prop} involved in the definitions of the query into
relations expressed in terms of Horn clauses, while the remaining inductive
types, including the browser state, the list of events, and the global
environment, are instead translated into SMT datatypes.
Note that existing tools for automation in Coq such as CoqHammer~\cite{Czajka2018,Czajka2020} and SMTCoq~\cite{ArmandFGKTW11} do not satisfy our needs: indeed, despite both relying on SMT solvers,
they focus
respectively on proof reconstruction and constraint solving, but not on
inhabitant finding.
We refer the interested reader to
Appendix~\ref{sec:appendix_compiler} for a discussion of the fragment of the
Coq logic supported by our compiler and for further details about the
compilation pipeline.

For every Web invariant that we aim to verify in our model, we define a corresponding query as a Coq
inductive type that is satisfied if a counterexample to
the invariant is found in any of the states reachable from the initial browser
state.
For example, let us consider the following invariant:
\begin{invariant}{\hspace{-4.2pt}}
  Cookies with the \code{Secure} attribute can only be set (using the
  \code{Set-Cookie} header) over secure channels.
\end{invariant}
\noindent
We encode this invariant in \theframework\ as follows:
\begin{lstlisting}[language=coq]
Definition SecureCookiesInvariant (gb: Global) (evs: list Event) (st: State) : Prop :=
  forall rp corr _evs cookie,
    Reachable gb evs st ->
    evs = (EvResponse rp corr :: _evs) ->
    rp_hd_set_cookie (rp_headers rp) = Some cookie ->
    sc_secure cookie = true ->
    url_protocol (rp_url rp) = ProtocolHTTPS.
\end{lstlisting}
This definition says that for every reachable state where the
browser  handles a network response, \ie, the state is \coqe{Reachable} and
the current event is \coqe{EvResponse} (lines 2-4), if the response contains a
\code{Set-Cookie} header (line 5) with a cookie that has the \code{Secure}
attribute (lines 6), then the protocol  used to serve the response is
HTTPS (line 7).

We encode a query for finding a counterexample to this invariant with the
following Coq inductive type:
\begin{lstlisting}[language=coq]
Inductive SecureCookiesQuery (gb: Global) (evs: list Event) (st: State) : Prop :=
| Query_state : forall rp corr _evs cookie,
    Reachable gb evs st ->
    evs = (EvResponse rp corr :: _evs) ->
    rp_hd_set_cookie (rp_headers rp) = Some cookie ->
    sc_secure cookie = true ->
    url_protocol (rp_url rp) <> ProtocolHTTPS ->
    SecureCookiesQuery gb evs st.
\end{lstlisting}
This inductive type definition is essentially identical to
\coqe{SecureCookieInvariant}, except the negation of the conclusion (line 7, we
require the protocol to be \coqe{<>} HTTPS).
The following theorem formalizes that inhabitants of \coqe{SecureCookiesQuery}
are indeed counterexamples of \coqe{SecureCookiesInvariant}:
\begin{lstlisting}[language=coq]
Theorem secure_cookies_query_invalidates_invariant :
  forall gb evs st, SecureCookiesQuery gb evs st ->
    not (SecureCookiesInvariant gb evs st).
\end{lstlisting}

\subsection{SMT Solving and Trace Reconstruction}
\label{sec_toolchain_solver}

The compilation of the inductive \coqe{Reachable} relation results in a
recursive CHC formula, which cannot be handled by standard SMT solvers.
Therefore we use the \muz\ extension (satisfiability modulo least fixed-points \cite{HBM11}) of the Z3 theorem prover. %
More precisely, we use in parallel the \spacer\ engine of \muz, a generalized property-directed reachability (GPDR) model checker suitable for finding proofs~\cite{HB12},
and the bounded model checking (\bmc)\ engine of \muz, designed to find counterexamples.
The four possible outcomes are: 
\begin{description}[font=\normalfont\itshape\bfseries, leftmargin=0pt]

  \item[\textsc{Sat}] \muz\ finds a counterexample, hence the invariant does not hold.
    We discuss in \autoref{sec:webinvariants} the security implications of
    violating an invariant.

  \item[\textsc{Unknown}] \muz\ fails to find a counterexample or to prove its absence.
    In such a case, which never happened in our case studies, we cannot draw
    any conclusion. 

  \item[\textsc{Unsat}] \muz\ proves that there is no counterexample.
    Although this does not formally suffice to conclude that the invariant holds in our model since neither
    our compiler nor \muz\ are formally verified, this gives us strong
    confidence that this is the case.
    A formal proof in Coq can be manually produced, if stronger confidence is
    needed.

  \item[\textsc{Loop}] \muz\ does not terminate.
    Due to the way the BMC engine works, this means that \muz\ did not find a counterexample
    after exploring a certain number of steps.
    When this number becomes high enough, though it is not a proof, it
    gives us a good intuition that the invariant is likely to hold,
    and hints it is worth starting a formal proof, as shown in
    \autoref{sec:proof}.

\end{description}

When running \theframework\ on \coqe{SecureCookiesQuery}, \spacer\ proves that
there is no counterexample within 2min, while in the same time, BMC reaches a
trace size of 50 events without detecting any attack.
Moreover, we  constructed a formal proof in Coq that the invariant
indeed holds~\cite{WebSpecSource}.

In the case \muz\ finds a counterexample (\textsc{Sat}), we first
automatically extract an attack trace from it.
It is worth noting that 
this attack trace enjoys the property of being minimal.
This property is due to the decision procedure implemented in the BMC engine of
\muz, which ensures that the list of events in the counterexample is the
smallest one that leads to a contradiction of the invariant.
Then, we verify its correctness by automatically translating it back into a
Coq term and checking whether the trace produces an inhabitant of the query.
We take this precaution because, as mentioned previously, neither our compiler
nor \muz\ are formally verified.
Since \muz\ instantiates all symbolic variables, the proof is straightforward
and mostly automatic.
\theframework\ then renders this trace as a sequence diagram, making the
representation of the counterexample accessible to users unfamiliar with
formal verification.
Examples of such diagrams are given in \autoref{sec:webinvariants}.

\subsection{Trace Validation}
\label{sec_toolchain_verifier}

In addition to rendering sequence diagrams, \theframework{} includes a verifier to validate the discovered attack traces against real-world browser implementations (Chrome, Firefox). Our verifier consists of approximately 3500 lines of OCaml code and leverages the Web Platform Tests (WPT)~\cite{WPT} cross-browser framework to map attack traces to tests, enabling verification against all major browsers. WPT is the standard test suite for browser and specification developers. It allows browser vendors to write tests modeling the expected behavior of Web standards and test their implementations for compliance. Using the common format specified by the WPT framework makes \framework{} tests compatible with the test suite. This allows for easy triage of the attack traces extracted from  counterexamples and the inclusion of our cross-component tests into WPT.

The generated tests translate trace events to browser actions and server responses. These actions modify the browser state to match the model's state after a given event. To maintain the browser and model states consistent throughout test execution, the verifier ensures these actions execute in the correct order. The effects of these actions are collected (directly or indirectly) and verified via WPT assertions. If all assertions succeed, the test is \emph{passing} and the trace is considered valid, whereas the test is \emph{failing} if one assertion fails.
We map each event to a $<$\emph{Setup}, \emph{Action}, \emph{Verification}$>$ (SAV) tuple:
\begin{description}[style=unboxed,leftmargin=0cm]
  \item[Setup]
    The setup is the set of pre-conditions necessary to execute an event, e.g., for \coqe{EvWorkerCacheMatch} to happen, a service worker must be installed in the matched URL's scope.
  \item[Action]
    An action is any browser or server action, from top-level navigations and JavaScript API calls to server responses. Actions can be implicit or explicit: \emph{Explicit} actions require an explicit API call or client-originated event, like window navigation; \emph{Implicit} actions are actions triggered by other actions (implicit or explicit), like subresource loading.
  \item[Verification]
    Verifications are the means to verify that an \emph{action} succeeded and can also be implicit or explicit. \emph{Explicit} verifications map to WPT assertions and are the primary mechanism for verifying attack traces. Asserting the value of a cookie after an \coqe{EvScriptSetCookie} event is an example of an explicit verification. \emph{Implicit} verifications do not require a WPT assertion to be verified. Server responses are an example of actions with implicit verifications, as they are not browser-dependent behavior and are known before runtime.
\end{description}

By mapping each event to an SAV tuple, we can generate an executable test with all the necessary steps to perform and verify each event in a trace. Since a \emph{passing} test verifies every event successfully, and each event evolves the browser's state to match the model's state, enforcing that events occur in the correct order implies that the browser's state at the end of a \emph{passing} test matches the model's state at the end of a trace, violating a given Web Invariant.

The following three paragraphs discuss some non-trivial implementation details necessary for trace verification:

\begin{description}[style=unboxed,leftmargin=0cm]
\item[Script Construction and Serialization.]
In our model, an \code{EvScript*} event is associated with a \code{DOMPath}~(\autoref{fig:dompath}) which identifies the \code{script} element performing the \emph{action} in the current window's DOM. The verifier keeps an ordered list of \code{EvScript*} events for each script element in a trace. Since each \code{EvScript*} event maps to a specific sequence of JavaScript methods, these events can be serialized. This serialization is also performed in order, adding verification and synchronization code between event actions when necessary.

\item[Event Synchronization.]
To keep tests as close as possible to the attack traces, the verifier must ensure that the executable test performs the trace events in the correct order. To enforce this property, we implemented token passing, meaning an event only executes if it holds a token. Without synchronization, scripts running in parallel on different pages, for example, could perform actions out of order, leading to inconsistent test results. For instance, take a trace where a script caches the response to a request subsequently matched by a service worker; we must ensure the caching occurs before the service worker match. Otherwise, the test no longer reflects the trace.

\item[Content-Security-Policy Inference.]
Invariants regarding the CSP impose a challenge on verification as browsers do not allow direct access to the CSP via JavaScript. Therefore, to verify the value of the CSP of a given browsing context, we must infer it. The verifier generates a set of URLs that the CSP should allow or block and calculates a signature representing this allow/block pattern. Pages updated via a \coqe{EvScriptDOMUpdate} then attempt to load scripts from these URLs. The verifier must also add a \code{report-uri} field to the CSP for the server component to know both allowed and blocked requests. The server can then calculate the CSP signature and report it to the main page, which asserts its value. If it matches, we conclude that the CSP matches the one in the trace.
\end{description}

\subsection{Discussion}
\label{sec_toolchain_discussion}

The choice of Coq for the formalization of a Web browser brings two advantages:
First, using
a strictly-typed higher-order language
as a specification language makes it possible to write expressive and
parametrizable models which can easily and consistently be extended to new Web
features.
Second, having our model specified within a proof assistant allows us to write
fully machine-checked proofs when the highest level of confidence is required.
We developed such proof for the four changes we propose to fix the
vulnerabilities that are currently affecting the Web platform.

In general, the main drawback of a model written in Coq is the lack of automation, which
becomes particularly problematic in the case of the model is  constantly evolving, like in our case to model the regular updates of the Web platform. 
Compiling our Web browser model from Coq to SMT queries circumvents this issue
by providing not only automatic counterexample finding as in
\cite{AkhaweBLMS10}, but also automated proofs that counterexamples do not exist.

Finally, a common limitation of model-based security analysis is the
lack of evidence of compliance between models and reality.  
In \theframework, we avoid this pitfall by using of a verifier (\autoref{sec_toolchain_verifier}) which allows us to execute and validate the discovered attack traces against real browsers.
In particular, we are able to validate the five vulnerabilities found in
our model of the current Web platform by running the attack traces produced by
\theframework\ against the latest versions of Chrome and Firefox.
Additionally, the verifier enables automated testing of the semantic rules of the browser model, ensuring that our model matches the observable behavior of real browsers. For every Web component in our model, we query for a reachable state which makes use of the modeled feature. Then we validate, using the verifier, that the obtained state maps to a reachable concrete state across browser implementations. Although this does not correspond to a correctness proof, it provides empirical evidence that our modeling is consistent with browser behavior.

\section{Web Invariants and Attacks}
\label{sec:webinvariants}

We define 10 Web invariants concerning the security properties of 5 core Web components: cookies, CSP, Origin header, SOP, and CORS. \autoref{tab:webinvariants} presents an overview of the invariants that we formally encode in our model. For each invariant that does not hold in the current Web platform, \framework{} is able to find a counterexample that translates to a concrete attack. When the invariant holds, \framework{} can be configured to reflect a past state of the Web that was affected by a vulnerability, confirming that our approach can identify previously reported attacks. 
In this section, we focus on three invariants that do not hold in the current Web platform, showing how \theframework\ is able to discover a new  attack on the \code{\_\_Host-} prefix for cookies as well as a new inconsistency between the Content Security Policy and a planned change in the HTML standard. We also present an attack against Trusted Types for which we propose a mitigation in \autoref{sec:proof}. Due to space constraints, we illustrate the encoding of the two remaining invariants that do not hold in the current Web  in \autoref{sec:appendix1}, while we present the full set of invariants in the technical report  \cite{webspec-technical-report}.

\begin{table*}[t]
  \scriptsize \centering
  \caption{Web Invariants}\label{tab:webinvariants}
  \begin{tabularx}{\linewidth}{llXp{7.4cm}cr}
    \toprule
    \textbf{Web Feature} & \multicolumn{2}{l}{\textbf{Invariant}} & \textbf{Description} & \textbf{Holds} & \textbf{References}\\
    \midrule
    \multirow{4}{*}{\textbf{Cookies}} & \textbf{I.1} & Integrity of \texttt{Secure} cookies (network) & \parbox{7.4cm}{Cookies with the \texttt{Secure} attribute can only be set (using the \texttt{Set-Cookie} header) over secure channels.} & \(\fullcirc\) & \cite{RFC6265bis}\\
    & \cellcolor{gray!20}\textbf{I.2} & \cellcolor{gray!20}Confidentiality of \texttt{HttpOnly} cookies (Web) & \cellcolor{gray!20}\parbox{7.4cm}{Scripts can only access the cookies without the \texttt{HttpOnly} attribute.} & \cellcolor{gray!20}\(\fullcirc\) & \cellcolor{gray!20}\cite{SinghMWL10,RFC6265}\\
    & \textbf{I.3} & Integrity of \texttt{\_\_Host-} cookies & \parbox{7.4cm}{A \texttt{\_\_Host-} cookie set for the domain $d$ can be set either by $d$ (via HTTP headers) or by scripts included by the pages on $d$.} & \(\emptycirc\) & \cite{SinghMWL10,CookiePrefixes}\\
    \hline
    \multirow{8}{*}{\textbf{CSP}} & \cellcolor{gray!20}\textbf{I.4} & \cellcolor{gray!20}Interaction with SOP & \cellcolor{gray!20}\parbox{7.4cm}{The DOM of a page protected by CSP can be read/modified only by the scripts allowed by the policy.} & \cellcolor{gray!20}\(\emptycirc\) & \cellcolor{gray!20}\cite{CSP,SomeBR17}\\
    & \textbf{I.5} & Integrity of server-provided policies & \parbox{7.4cm}{If a response from the server contains a security policy, then the browser enforces that specific policy.} & \(\emptycirc\) & \cite{SquarcinaCM21}\\
    & \cellcolor{gray!20}\textbf{I.6} & \cellcolor{gray!20}Access control on Trusted Types DOM sinks & \cellcolor{gray!20}\parbox{7.4cm}{If a page has both \texttt{trusted-types;} and \texttt{\hyphenchar\font=`\-require-trusted-types-for 'script';} directives in the CSP then no script in the page can modify the DOM using a Trusted Types sink.} & \cellcolor{gray!20}\(\emptycirc\) & \cellcolor{gray!20}\cite{TrustedTypesDraft}\\
    & \textbf{I.7} & Safe policy inheritance & \parbox{7.4cm}{Documents loaded from a local scheme inherit the policy of the source browsing context.} & \(\halfcirc\) & \cite{CSP}\\
    \hline
    \multirow{1}{*}{\textbf{Origin Header}} & \cellcolor{gray!20}\textbf{I.8} & \cellcolor{gray!20}Authenticity of request initiator & \cellcolor{gray!20}\parbox{7.4cm}{If a request $r$ includes the header \texttt{Origin:~$o$} (with $o \neq \texttt{null}$), then $r$ was generated by origin $o$.} & \cellcolor{gray!20}\(\fullcirc\) & \cellcolor{gray!20}\cite{AkhaweBLMS10,W3CCORS}\\
    \hline
    \multirow{3}{*}{\textbf{SOP/CORS}} & \textbf{I.9} & Authorization of non-simple requests ($i$) & \parbox{7.4cm}{A non-simple cross-origin request must be preceded by a pre-flight request.} & \(\fullcirc\) & \cite{AkhaweBLMS10,W3CCORS09}\\
    & \cellcolor{gray!20}\textbf{I.10} & \cellcolor{gray!20}Authorization of non-simple requests ($ii$) & \cellcolor{gray!20}\parbox{7.4cm}{The authorization to perform a non-simple request towards a certain origin $o$ should come from $o$ itself.} & \cellcolor{gray!20}\(\fullcirc\) & \cellcolor{gray!20}\cite{AkhaweBLMS10,HTML}\\
    \bottomrule
  \end{tabularx}
  \ \\ \ \\
  {\scriptsize
    \(\halfcirc\) The invariant holds in the current version of the Web platform but a planned modification will invalidate it.
  }
\end{table*}

\subsection{Integrity of \texttt{\_\_Host-} Cookies}
\label{sec:hostdomainrelaxation}

The invariant stipulates that \texttt{\_\_Host-} cookies ensure integrity against same-site attackers \cite{SquarcinaTVCM21}. When a cookie whose name starts with \texttt{\_\_Host-} is set, the browser verifies that the \texttt{Domain} attribute is not present and discards the cookie otherwise, thus marking all \texttt{\_\_Host-} cookie as \texttt{host-only}.
\begin{invariant}{I.3}
  A \texttt{\_\_Host-} cookie set for the domain $d$ can be set either by $d$ (via HTTP headers) or by scripts included by the pages on $d$.
\end{invariant}

We encode this invariant  by splitting the two cases in which a host cookie can be set: (i) via HTTP headers, and (ii) via JavaScript.
For space reasons we present only case (ii) below and refer to the technical report~\cite{webspec-technical-report} for the full definition.
\begin{lstlisting}[language=coq]
Definition HostInvariantSC (gb: Global) (evs: list Event) (st: State) : Prop :=
  forall pt sc ctx c_idx cookie cname h _evs,
    Reachable gb evs st ->
    (* A script is setting a cookie *)
    is_script_in_dom_path gb (st_window st) pt sc ctx ->
    evs = (EvScriptSetCookie pt (DOMPath [] DOMTopLevel) c_idx cookie :: _evs) ->
    (* The cookie prefix is __Host *)
    (sc_name cookie) = (Host cname) ->
    (* The cookie has been set in the script context *)
    url_host (wd_location ctx) = Some h ->
    (sc_reg_domain cookie) = h.
\end{lstlisting}
For every reachable state in which a script \coqe{sc} is setting a cookie on the top-level window (lines 3-6), \coqe{ctx} is the window (browsing context) in which the script \coqe{sc} is running (line 5).
If the cookie has the \code{\_\_Host-} prefix (line 8), we require (line 11) the domain on which the cookie was registered to be equivalent to the domain of the \coqe{ctx} browsing context.
This corresponds to stating that a script running on a page of domain $d$ can set a host-prefix cookie only for the domain $d$.

\bfparagraph{Attack.}
When we run the query, \thetool\ discovers a novel attack
that breaks the invariant using domain relaxation.
A script running on a page can modify at runtime the \emph{effective} domain used for SOP checks through the \texttt{document.domain} API.
Indeed, the value of \texttt{document.domain} is taken into account only for DOM access. All remaining access control policies implemented in the browser use the original domain value~\cite{SinghMWL10}. This is the case, for instance, for cookie jar access, XMLHttpRequests, and origin information reported when performing a \texttt{postMessage}.
The mismatch between the access control policies in the DOM and the cookie jar allows a script running in an iframe to access the \texttt{document.cookie} property of the parent page when both pages set \texttt{document.domain} to the same value. Once the inner frame performs a set cookie of a host-prefix cookie through the parent page DOM, the browser uses the original domain value of the parent page to perform the host prefix checks, breaking the invariant.

\begin{figure}[t]
  \centering
  \includegraphics[width=0.30\textwidth]{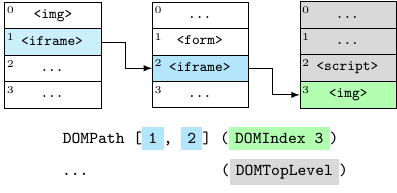}
  \caption{DOM Path Datatype}
  \label{fig:dompath}
\end{figure}

\begin{figure}[t]
  \centering
  \includegraphics[width=\linewidth]{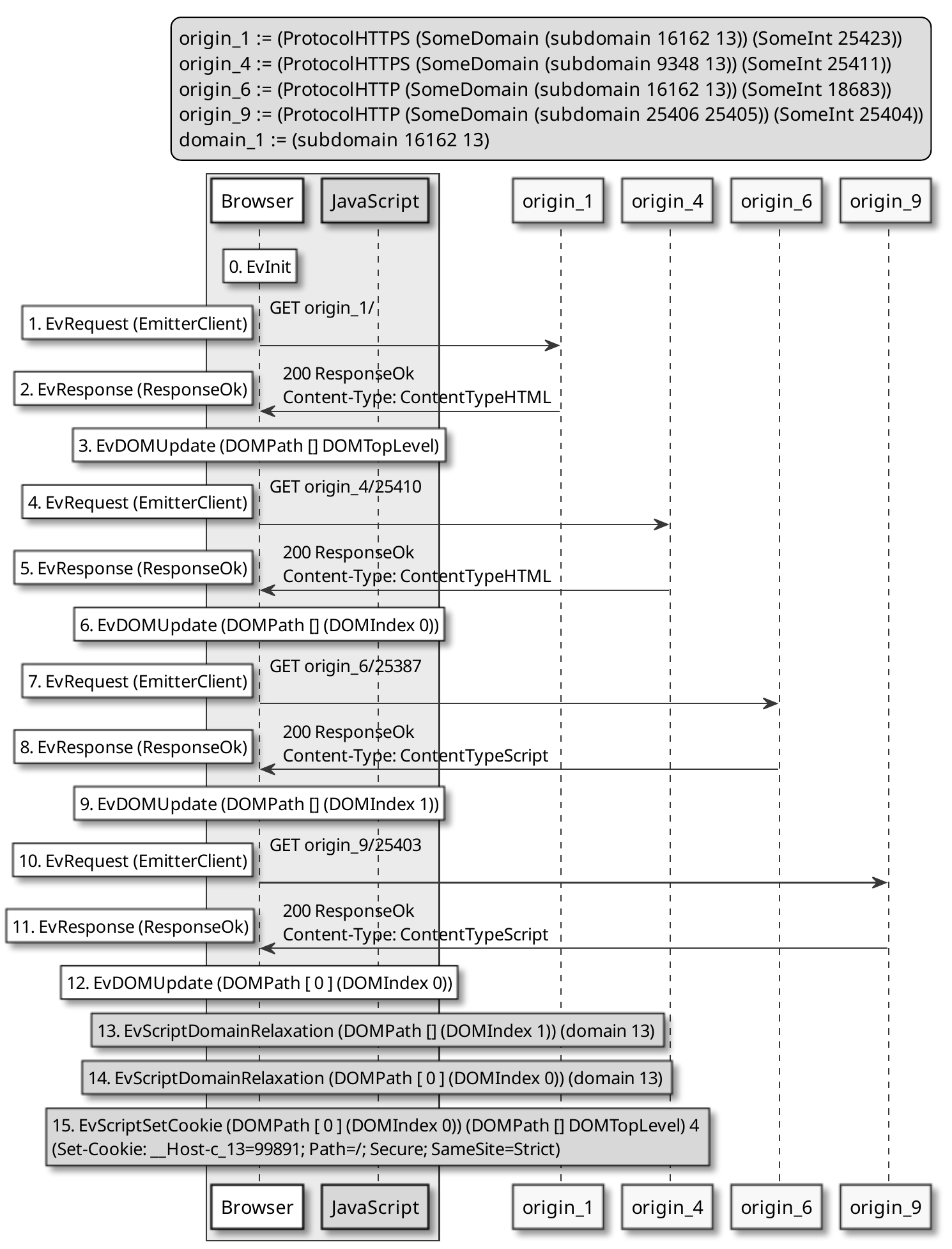}
  \caption{Host Cookies Inconsistency}
  \label{fig:hostcookies}
\end{figure}

The trace generated by \theframework\ is shown in \autoref{fig:hostcookies} and detailed below.
In the following, expressions of the form \coqe{DOMPath _ _}\ represent a unique path in the DOM.
In particular, the first argument of \coqe{DOMPath} is the nesting level. For instance, we refer to the window loaded inside two nested iframes as \coqe{DOMPath [1,3] _}, where 1 and 3 are the indexes of the DOM elements representing the frames. The second argument is used
to refer to a specific DOM object (\coqe{DOMIndex}) or to the whole document loaded in the frame (\coqe{DOMTopLevel}). An example is shown in \autoref{fig:dompath}: the path to an image at index 3 loaded inside two nested ifranes (respectively at index 2 and 1) is represented as \coqe{DOMPath [1,2] (DOMIndex 3)}, while the path of the window containing the image is \coqe{DOMPath [1,2] DOMTopLevel}.

The attack trace describes the following scenario:
\begin{enumerate*}
\item[(steps 1-3)] a page from \code{origin\_1} is loaded in the top-level window of the browser.
 Note that \code{origin\_1} is the subdomain named \code{16162} of the host \code{13}, loaded via HTTPS;
\item[(4-6)] an \code{iframe} element is loaded from \code{origin\_4} at index 0 of the DOM in the main window (in the path \coqe{DOMPath [] (DOMIndex 0)}).
\code{origin\_4} is another subdomain of the same host;
\item[(7-9)] a script is loaded in the main window at index 1;
\item[(10-12)] a script is loaded in the iframe at index 0 (\coqe{DOMPath [0] (DOMIndex 0)});
\item[(13)] the script in the parent window sets its \code{document.domain} to its parent domain \coqe{13};
\item[(14)] the script in the iframe sets its \code{document.domain} to its parent domain \coqe{13}. The two pages are now effectively same origin, having performed domain relaxation to the same domain;
\item[(15)] the script inside the iframe (\coqe{DOMPath [0] (DOMIndex 0)}) sets a cookie using \code{document.cookie} of the top-level window (\coqe{DOMPath [] DOMTopLevel}). The cookie has the \code{\_\_Host-} prefix and has been set by
  \code{origin\_1} for \code{origin\_2}, breaking the invariant.
\end{enumerate*}

Although the current Web platform is still vulnerable to the attack, discontinuing the \code{document.domain} API will eventually make the invariant hold.
\theframework{} can reflect this change by specifying \coqe{c_domain_relaxation (config gb) = false}, allowing us to verify
that the invariant holds. 

\subsection{Access control on Trusted Types DOM sinks}
\label{sec:trustedtypes}

Trusted Types allow for \emph{locking down} a document by disabling DOM XSS sinks entirely. This special setting corresponds to the following invariant.

\begin{invariant}{I.6}
  If a page has both \texttt{trusted-types;} and \texttt{require-trusted-types-for} \texttt{'script';} directives in the CSP then no script in the page can modify the DOM using a Trusted Types sink.
\end{invariant}

\noindent We encode the invariant in \themodel\ as follows:
\begin{lstlisting}[language=coq]
Definition TTInvariant (gb: Global) (evs: list Event) (st: State) : Prop :=
  forall pt target_pt target_ctx ssrc ttypes,
    Reachable gb evs st ->
    (* The target context has Trusted-Types enabled *)
    url_protocol (wd_location target_ctx) = ProtocolHTTPS ->
    rp_hd_csp (dc_headers (wd_document target_ctx)) = Some
      {| csp_script_src := ssrc; csp_trusted_types := Some ttypes |} ->
    tt_policy ttypes = Some None ->
    tt_require_for_script ttypes = true ->
    (* No script can update the dom using innerHTML *)
    not (In (EvScriptUpdateHTML pt target_pt target_ctx) evs).
\end{lstlisting}
Here we assert that there cannot be an html update event (using a DOM XSS sink, \eg, \coqe{innerHTML})
for the window \coqe{target_ctx}, if the aforementioned directives are used to define the policy for \coqe{target_ctx}.

\bfparagraph{Attack.}
An earlier version of the Trusted Types draft \cite[Editor's Draft, Feb. 3, 2021]{TrustedTypesDraft} restricted Trusted Types to Secure Contexts only.
This was part of an effort of browser vendors to restrict all new APIs to secure contexts to help advance the Web platform to default to the HTTPS protocol.
The restriction, however, enabled attackers to bypass Trusted Types by framing the protected page from a non-secure context~\cite{TTSecureCtx}.
This silently disabled the DOM XSS protection despite the fact that the document was downloaded using a secure connection.
When we enable the secure context restriction in \themodel, \theframework\ is able to rediscover the bypass.

\begin{figure}[t]
  \centering
  \includegraphics[width=0.43\textwidth]{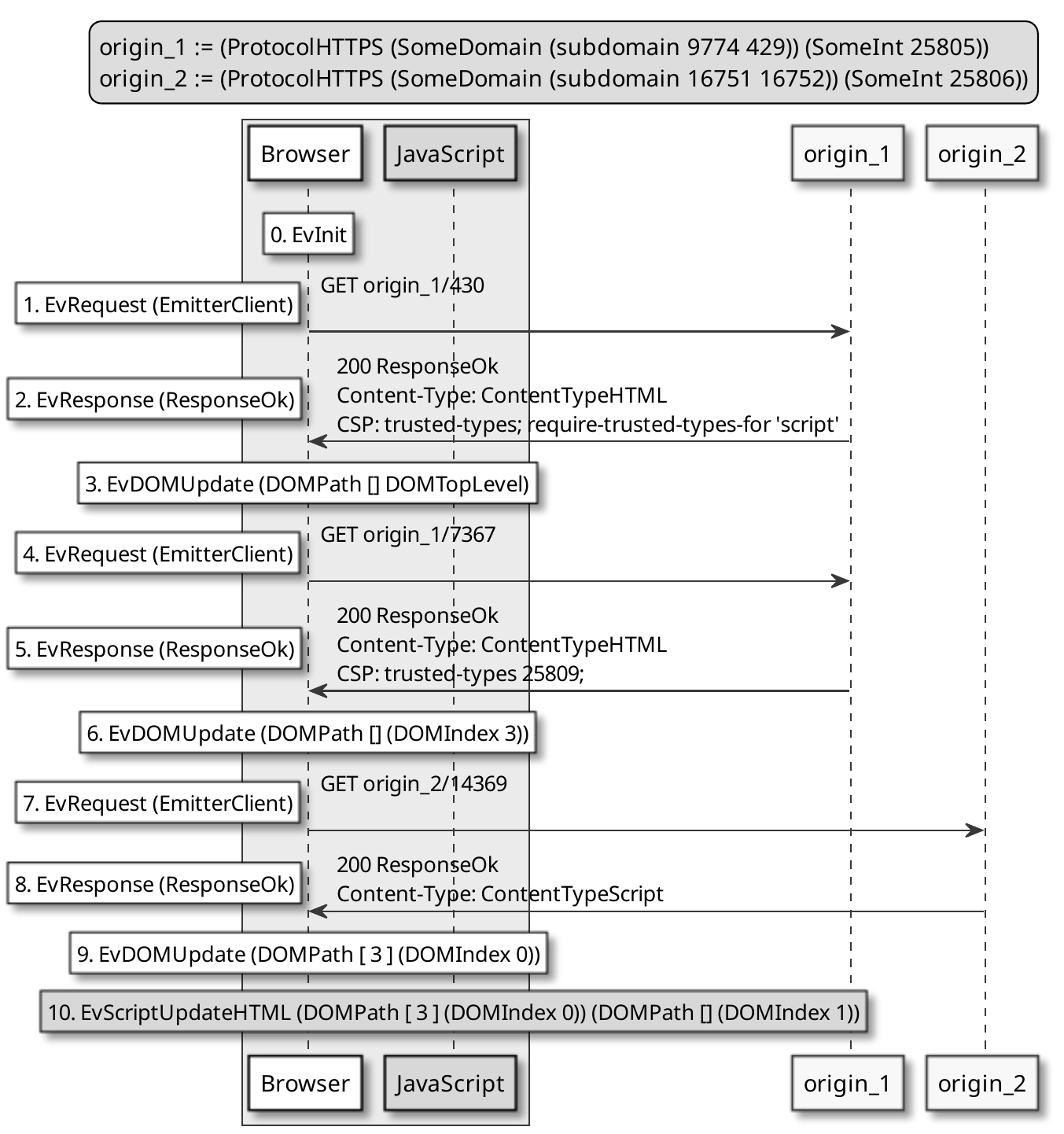}
  \caption{Trusted-types bypass with same-origin iframes}
  \label{fig:ttcolluding}
\end{figure}

We can disable the secure context restriction with the
\coqe{c_restrict_tt_to_secure_contexts (config gb) = false} configuration option.
However, when we rerun the solver with this configuration, \thetool\ can still find a counterexample for which the invariant does not hold.
The trace is shown in \autoref{fig:ttcolluding}:
\begin{enumerate*}[label=(\roman*)]
\item[(steps 1-3)] a page protected with Trusted Types is loaded from \texttt{origin\_1}. In particular, no policy is allowed, so no Trusted Type can be created;
\item[(4-6)] the page contains a same-origin iframe which specifies a Trusted Types policy (\texttt{trusted-types 25809}), allowing scripts loaded in this iframe to create Trusted Types using a policy named \texttt{25809};
\item[(7-9)] a script that is loaded in the iframe modifies (10) the DOM of the parent frame using a Trusted Types sink. This is possible because the inner frame is able to create Trusted Types that are accepted by all DOM XSS sinks and because, being same origin, the inner frame can access the DOM of the parent.
\end{enumerate*}
A similar attack on related domains is possible if the parent page performs domain relaxation, as the value of \texttt{document.domain} is used for DOM access control.

The Trusted Types draft~\cite[§5.1]{TrustedTypesDraft} includes a brief discussion of a similar attack in which cross-document import of nodes would bypass the enforcement of the policy. However, the current specification does not provide any solution and suggests that other mechanisms like \emph{Origin Policy}~\cite{OriginPolicyDraft} might be used to ensure that the same policy is deployed across the whole origin.
Instead, we propose a different solution based on \emph{non-transferable} Trusted Types, and prove the correctness of our approach within our model in \autoref{sec:proof}.

\subsection{Safe policy inheritance}
\label{sec:csplocalschemes}

The Content Security Policy specification \cite[§7.8]{CSP} mandates that every document loaded from a local scheme
must inherit a copy of the policies of the source browsing context, \ie, the browsing context that was responsible for starting the navigation.
This corresponds to the following invariant.
\begin{invariant}{I.7}
  Documents loaded from a local scheme inherit the policy of the source browsing context.
\end{invariant}

\noindent We encode the invariant in \themodel\ as follows:
\begin{lstlisting}[language=coq]
Definition LSInvariant (gb: Global) (evs: list Event) (st: State) : Prop :=
  forall evs pt _evs frm fhtml fwd ctx lv pt_idx init_idx,
    let get_csp wd :=
        rp_hd_csp (dc_headers (wd_document wd)) in
    Reachable gb evs st ->
    (* A document has just been loaded in a frame *)
    evs = (EvDOMUpdate pt :: _evs) ->
    is_frame_in_dom_path gb (st_window st) pt frm fhtml fwd ctx ->
    is_local_scheme (wd_location fwd) ->
    (* get navigation initiator *)
    pt = DOMPath lv (DOMIndex pt_idx) ->
    is_wd_initiator_of_idx ctx pt_idx (Some init_idx) ->
    (* The csp is equal to the req. initiator *)
    get_csp fwd = get_csp (windows gb.[init_idx]).
\end{lstlisting}
When a frame has just loaded a document from a local scheme (lines 7-9), we require that the CSP of the navigation initiator (\ie, the source browsing context) is equal to the policy of the document loaded in the frame window (line 14).

The goal of this Web invariant is to ensure that a page cannot bypass its policy by navigating to content that is completely under its control. One such bypasses~\cite{ChromiumBugLSCSPBypass} was caused by the behavior defined for the inheritance of policies in a previous version of the CSP specification \cite[Oct. 15, 2018]{CSP}: documents loaded from local schemes would inherit the policies of the embedding document or the opener browsing context.

Recently, the concept of \emph{policy container} was added to the HTML specification \cite[§7.9]{HTML}. A policy container is a collection of policies to be applied to a specific document and its purpose is to simplify the initialization and inheritance of policies.
The introduction of the policy container in the specification allowed for clarifying the inheritance behavior for local schemes, which might differ depending on the specific scheme or URL that is used.  The policy container explainer~\cite{PolicyContainerExplainer} stipulates the following behavior:
\begin{description}[leftmargin=0pt]
\item[\emph{about:srcdoc}] An \texttt{iframe} element with the \texttt{srcdoc} attribute inherits the policies from the embedding document, \ie, the parent frame. Note that srcdoc iframes are in the same origin of the embedding document but their location URL is \texttt{about:srcdoc}.
\item[\emph{about:}, \emph{data:}] A document loaded from the \texttt{data:} or \texttt{about:} schemes inherits the policies of the navigator initiator (as mandated by the CSP specification).
\item[\emph{blob:}] A document loaded from a \texttt{blob:} URL inherits the policies from the document that creates the URL, \ie, the document that calls the \texttt{URL.createObjectURL} function.
\end{description}

Note that in the current version of the HTML specification \cite[§7.11.1]{HTML} the inheritance behavior for \texttt{blob:} URLs matches the one for \texttt{about:} and \texttt{data:}, thus following the CSP specification.
We contacted the editors of the HTML specification~\cite{OurLSIssue}
asking for a clarification on the correct behavior for \texttt{blob:} URL  and they confirmed that, because of the wrong ordering of a clause in the policy container construction for blobs, the initiator policy container was always replacing the creator policy container.
The correct inheritance rule is to inherit the policy container of the creator of the URL~\cite{LSPullRequest}, thus introducing an inconsistency between the CSP specification and the HTML specification (as \texttt{blob:} is a local scheme that is handled differently from the others).

\begin{figure}[t]
  \centering
  \includegraphics[width=0.45\textwidth]{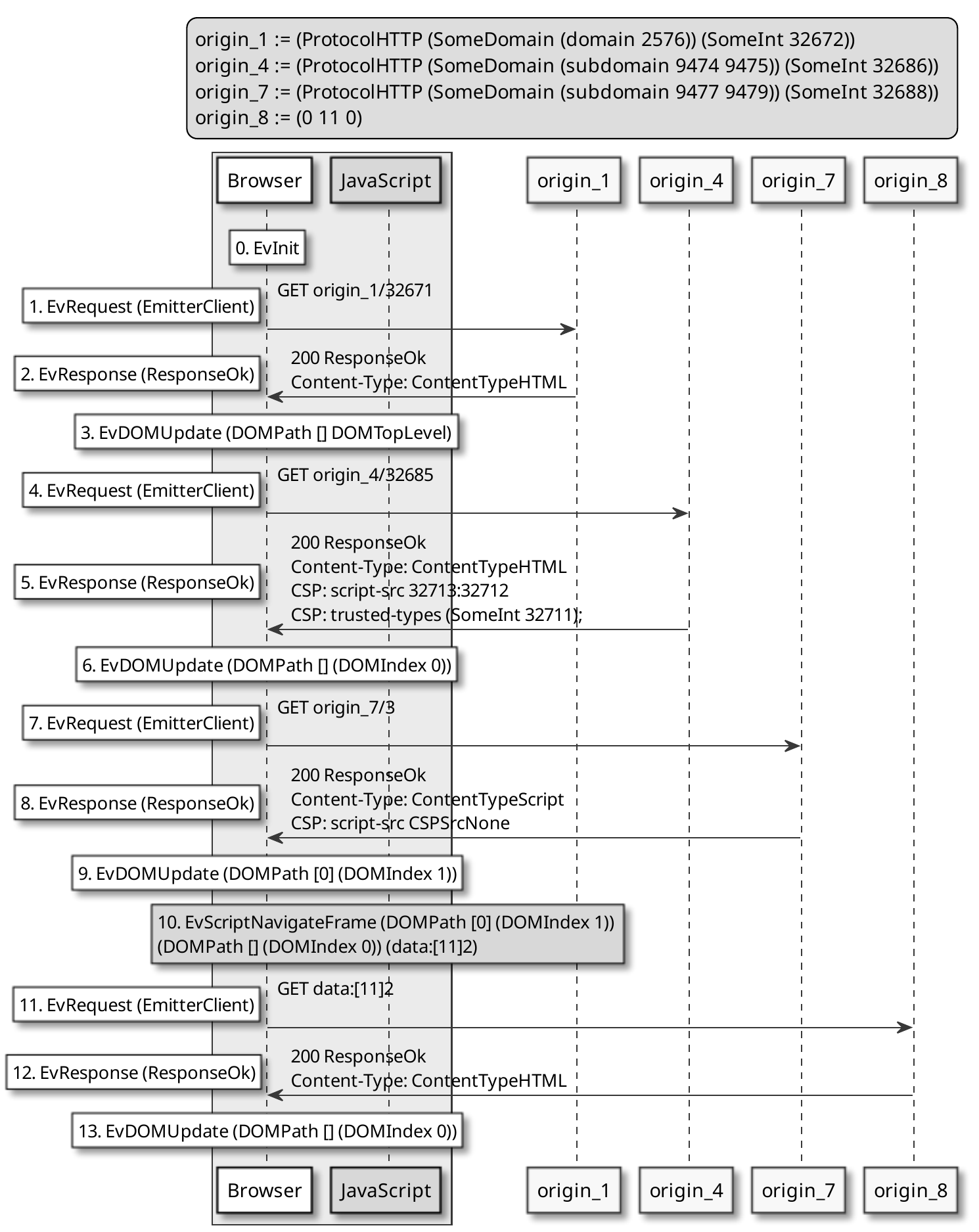}
  \caption{CSP bypass due to inheritance from the embedder document}
  \label{fig:inheritanceparent}
\end{figure}

\bfparagraph{Attack.}
When we configure \themodel\ to reflect a past state of the Web platform in which policies were inherited from the embedding frame and not from the navigation initiator
(\coqe{c_csp_inherit_local_from_initiator (config gb) = false}), \thetool\ is able to rediscover the attack trace that allows an attacker to strip the CSP policies by navigating a frame to a local scheme URL.
The trace is shown in \autoref{fig:inheritanceparent}:
\begin{enumerate*}[label=(\roman*)]
\item[(steps 1-6)] a document with no Content Security policy loads an iframe with a restrictive CSP;
\item[(7-9)] the iframe contains a script which navigates (10) the frame itself (\eg, using the \texttt{window.location} setter) to a local scheme URL;
\item[(11-13)] the iframe renders the content of the local scheme URL and inherits the CSP from the embedding document, which contains no policy. The resulting document has no CSP, effectively removing the iframe's previous policy.
\end{enumerate*}

We can configure \themodel\ to reflect the current state of the Web platform by inheriting the policies from the navigation initiator for all local schemes: \begin{lstlisting}[language=coq,numbers=none]
c_csp_inherit_local_from_initiator (config gb) = true /\
c_csp_inherit_blob_from_creator (config gb) = false.
\end{lstlisting}%
We can verify (up to a finite size, see \autoref{sec:evaluation}) that with this configuration the invariant holds.
However, when we configure \themodel\ to reflect the planned modification of inheriting the policies of the URL creator when rendering a \texttt{blob:} URL (\coqe{c_csp_inherit_blob_from_creator (config gb) = true}), Z3 is able to find a new counterexample. The trace is similar to the one depicted in \autoref{fig:inheritanceparent} with an additional script loaded in the top-level window executing an \coqe{EvScriptCreateBlobUrl} event: 
\begin{enumerate*}[label=(\roman*)]
\item a page in \texttt{origin\_1} with no CSP loads a same-origin iframe with a restrictive policy;
\item a script running on the embedding document creates a new blob URL (\coqe{EvScriptCreateBlobUrl});
\item a script running on the inner frame navigates the frame itself (\ie, setting \code{window.location}) to the previously created URL. The frame loads the content of the blob and inherits the CSP from the embedding document, which does not have any policy. Similarly to the previous attack trace, the policy that was defined for the iframe has been removed by navigating to local-scheme content.
\end{enumerate*}

Hence, the planned modification on CSP inheritance in the HTML standard would introduce an inconsistency with the CSP specification. We responsibly reported the issue to the working group of the HTML standard~\cite{OurLSIssue}, who initially deemed the security implications of the attack as low. However, at the time of writing (August 2022), no final decision has been taken and the current browser behavior remains unchanged, \ie, our invariant still holds.

\section{Verification of Security Properties}
\label{sec:proof}
\framework{} supports both machine-checked and automated Web invariant proof generation. In particular, we wrote four machine-checked proofs in Coq and were able to automatically derive two further Web invariant proofs through the Spacer engine of $\mu$Z. Due to space constraints, we  show here how to formally verify the security of the  fix we present against the attack on Trusted Types (\autoref{sec:trustedtypes}) through a machine-checked proof in \framework{}, referring to the long version~\cite{webspec-technical-report} and to the online repository~\cite{WebSpecSource} for the remaining proofs.

According to the current draft of the Trusted Types specification, a Trusted Type object created by a page can be assigned to DOM XSS sinks belonging to different pages.
This allows for bypassing the protection if a restricted document colludes with an unrestricted one. This can happen, \eg, in case of same-origin iframes (see \autoref{sec:trustedtypes}).
The specification acknowledges the issue and suggests the usage of the Origin Policy~\cite{OriginPolicyDraft} to address the problem, which unfortunately is not currently supported by any browser.

For this reason, we propose an alternative solution that we label \emph{non-transferable} Trusted Types, which consists in labeling each Trusted Type with the JavaScript realm (window or worker) that created it and ensuring that a type can only be assigned to DOM XSS sinks from the same realm. Therefore, our fix prevents the cross-document usage of Trusted Types.
We implemented this behavior in \framework{}, which can be activated by setting the configuration option \coqe{c_tt_strict_realm_check} to \coqe{true}.
The following theorem states the validity of the invariant \texttt{TTInvariant} (\autoref{sec:trustedtypes}) when our fix is enabled:
\begin{lstlisting}[language=coq]
Theorem strict_realm_check_implies_invariant :
  forall gb evs st,
     c_tt_strict_realm_check (config gb) = true ->
     c_restrict_tt_to_secure_contexts (config gb) = false ->
     TTInvariant gb evs st.
\end{lstlisting}
We recall that, according to the invariant, if a page is shipped with a CSP containing the directives \texttt{trusted-types} and \texttt{require-trusted-types-for 'script'}, then the list of events \texttt{evs} cannot contain a \texttt{EvScriptUpdateHTML} event that updates the contents of the page.

We can prove the theorem by induction on the \texttt{Reachable} relation where all the cases except \texttt{EvScriptUpdateHTML} are trivial.
In this latter case we show that, by enabling strict realm checking, it is impossible to generate the correct Trusted Type for the update, since
\begin{enumerate*}[label=(\roman*)]
	\item the \texttt{trusted-types} directive disallows the creation of Trusted Types for the realm in which the directive is used; and
	\item the only Trusted Types that are accepted by a page with \texttt{require-trusted-types-for 'script'} are only those labelled with the realm of the page.
\end{enumerate*}
This suffices to prove the correctness of the proposed solution within our model. The whole proof in Coq spans just over 54  LOC. For comparison,  the longest of the four proofs we conducted in Coq is just 348 LOC, which demonstrates the feasibility of  writing machine-checked proofs in our model.

\section{Evaluation}
\label{sec:evaluation}

\begin{table*}[t]
\newcommand{\Cgrn}{\cellcolor{green!18}}
\newcommand{\tocheck}[1]{#1}
\newcommand{\mehh}{$^\star$} 
\centering \footnotesize
\caption{Trace size and solving time for each attack}\label{tab:evaluation}
\vspace{-0.2cm}
\begin{tabularx}{\linewidth}{rX|c|cccc}
\multicolumn{3}{c}{} & \multicolumn{4}{c}{\textbf{Solving Time}} \\
\toprule
\multirow{2}{*}{\textbf{\#}} & \multirow{2}{*}{\textbf{Query}} & \textbf{Trace Size} & \multicolumn{2}{c}{\textbf{Frames Enabled}}  &  \multicolumn{2}{c}{\textbf{Frames Disabled}}  \\
 &  & \textbf{\# Events} & \textbf{Baseline} & \textbf{w/ Lemmas} & \textbf{Baseline} & \textbf{w/ Lemmas} \\
\midrule
1 & Integrity of \texttt{\_\_Host-} cookies & 15 & 58d 1h 30m & \Cgrn 23m \mehh & \(\times\) & \(\times\)\\
2 & Confidentiality of \texttt{HttpOnly} cookies & 7 & 13h 35m & 8m & 1h 46m & \Cgrn 1m\\
3 & Interaction between SOP and CSP & 10 & 42d 3h & \Cgrn 46m \mehh & \(\times\) & \(\times\)\\
4 & Integrity of server-provided policies & 9 & 40h 35m & 18m & 15h 46m & \Cgrn 3m\\
5 & Access control on Trusted Types sinks & 9 & \Cgrn 17h 50m & -- & \(\times\) & \(\times\)\\
6 & Access control on Trusted Types sinks (no sec. ctx) & 10 & \Cgrn 9d 20h 18m & -- & \(\times\) & \(\times\)\\
7 & Safe policy inheritance (inherit from parent) & 13 & 52d 10h & \Cgrn 1h 48m & \(\times\) & \(\times\)\\
8 & Safe policy inheritance (inherit from creator) & 17 & -- & \Cgrn 6h 5m & \(\times\) & \(\times\)\\
9 & Authenticity of request Initiator & 5 & 1h 49m & -- & \Cgrn 14m & --\\
10 & Authorization of non-simple requests (\(i\)) & 5 & 35m & -- & \Cgrn 5m & --\\
11 & Authorization of non-simple requests (\(ii\)) & 10 & -- & 48m & 35d 10h 51m & \Cgrn 7m\\
\bottomrule
\end{tabularx}
\ \\ \ \\
{\scriptsize
  $\times$: N/A; \qquad
  -- (Baseline): No solution could be found within 60 days; \qquad
  -- (w/Lemmas): None of our user-defined lemmas could be applied; 
  \\ 
  $\star$: a lemma has been automatically extracted from the attack trace of a previous run of the solver.
}
\end{table*}

In our experimental evaluation, we used \framework{} to automatically discover the attacks reported in \autoref{tab:webinvariants}. Additionally, when we implemented a fix to an attack, we ran \framework{} again to confirm that the issue had been addressed. Since the $\mu$Z solver may not terminate (see \autoref{sec:compiler}), we use the length of the previously discovered attack trace plus one as the maximal search size, thus obtaining confidence that the previous attack is not reachable anymore.

We report the time required by \framework{} to find the attacks and describe various optimization techniques that allowed us to drastically improve the performance of our approach. All our experiments have been conducted on a virtual machine with 32 VCPUs (2GHz AMD EPYC) and 128GB of RAM.

The baseline performance is displayed on the third column of \autoref{tab:evaluation}.
We can observe a clear correlation between the size of the attack trace and the time required to find an attack, which is caused by the unrolling technique employed by the BMC engine of the $\mu$Z solver used in \framework.
In particular, time increases exponentially with respect to the size of the attack trace, leading to running times of several days or weeks for traces with 10 or more events.
To tackle these performance issues, we have implemented various optimizations that consist in
\begin{enumerate*}[label=(\roman*)]
\item defining additional rules (or lemmas) representing common configurations (e.g., loading of a frame containing a script) that can be used by the SMT solver instead letting it rediscover the right list of events leading to these configurations, and
\item simplifying the model at compile time (e.g., by disabling frames) so that the resulting SMT formula is easier to solve.
\end{enumerate*}

We describe these optimizations in the following, and we refer the reader to \autoref{sec:appendix-scalability} for a discussion on the scalability of our browser model, which confirms the effectiveness of lemmas in mitigating the complexity that arises from the addition of new Web components.

\subsection{User-defined Lemmas}
\label{sec:evaluation-lemmas}

The key idea underlying this optimization is to enable users to define additional lemmas that guide the solver into constructing interesting browser states that can be used as a starting point to discover new attacks.
Intuitively, lemmas represent encodings of common Web Security threat models, which map to the general preconditions of specific classes of Web attacks.
Consider the following example:
\begin{lstlisting}[language=coq]
Lemma script_state_is_reachable : forall gb,
    script_state_constraints gb ->
    Reachable gb (script_state_events gb) script_state.
\end{lstlisting}
Here \coqe{script_state} is a Coq definition of a concrete browser state where a script is loaded in the page rendered in the top-level window.
This state maps to the threat model in which an attacker controls a script running in the same page as the target Web application, as a result of, e.g., XSS or the inclusion of untrusted scripts \cite{NikiforakisIKAJKPV12}.
The lemma encodes that this browser state is reachable by applying the list of events \coqe{script_state_events} assuming that \coqe{script_state_constraints} is satisfied.
Once we prove that the state defined by a lemma is \coqe{Reachable}, the lemma can be compiled as an additional model rule to CHC logic.
Since the BMC engine solves queries by iterative unrolling, it prioritizes the rules that result in the smallest amount of unrolling steps. Lemmas exploit this property by providing a one-step solution for the generation of states that would require multiple steps if the solver had to build them from scratch.

\framework\ includes the definition of five generic lemmas, which represent two variants of the aforementioned attacker controlled script, one running in a secure context and the second being served from an insecure connection, and three variants of the \emph{gadget attacker} of \cite{BarthJM08B} in which a same-origin, same-site or cross-site iframe containing an attacker controlled script is included into the target Web applications. 
When compiling the model to an SMT formula, the \coqe{Using Lemmas} directive is used to specify which user-defined lemmas need to be considered for the query. For our experiments, we provided every query with all the lemmas that are included in \framework, since we experimentally observed that the number of lemmas added to a query does not negatively impact solving time. This behavior may result from the effectiveness of \muz\ in discarding all non applicable initial states.
For this reason, we expect only an improvement of the solver performance from the definition of a larger library of lemmas.

The results of this optimization are highlighted in \autoref{tab:evaluation}, where we can see that the usage of lemmas always reduces the runtime to less than a day.
It may however happen that the solver is not able to apply any of the user-defined lemmas, as it is the case for queries \#5 and \#6 (marked with --). In such cases no performance improvement can be obtained.
For the queries marked with~$\star$, we automatically extracted a lemma from an attack trace discovered by \theframework{} and confirmed that it can be used by successive runs of the solver.
The extraction of lemmas from traces has several benefits: on the one hand, it allows to quickly test for the absence of a known attack after applying a fix to the model; on the other hand, it allow us to add new reachable browser configurations to our library of lemmas. Since these configurations represent the preconditions for the execution of a specific attack, the extraction of additional lemmas augments our counter example pipeline with a method to quickly discover novel attacks assuming known preconditions.
We leave the extension of this library and the definition of a methodology to generate generic lemmas as future work.

\subsection{Compile-Time Simplification}

\bfparagraph{Configurable inlining of auxiliary relations.}
Our model relies on a \texttt{Reachable} relation that models state transitions, a \texttt{ScriptState} relation that models scripts knowledge, and multiple auxiliary relations that are used within \texttt{Reachable} to, \eg, recursively update the DOM.
The presence of multiple relations prevents us to directly use the best performing version of the BMC engine, the linear solver, because it requires the model to be encoded as linear Horn clauses, \ie, clauses containing at most one recursive term.
In order to satisfy this requirement, every auxiliary relation needs to be inlined within the main \texttt{Reachable} relation.
To this end, \framework{} automatically unrolls all the applications of recursive relations that are marked for inlining.
For each relation we specify the depth of the unrolling. For instance, the declaration
\begin{lstlisting}[language=coq,numbers=none]
Inline Relation is_script_in_dom_path With Depth 3.
\end{lstlisting}
says that the relation \coqe{is_script_in_dom_path}, which searches a script inside the DOM, must be unrolled up to recursion level 3.
Depth 0 disables all recursive calls and expands the relation to the base case only. For instance, support for nested frames can be easily deactivated by specifying 0 as depth for all the relations handling the DOM tree.

The recursion depth affects the solving time of $\mu$Z since multiple applications of the relation need to be considered. Disabling nested frames for the queries which do not require them simplifies the compiled model and allows for faster solving. When frames are required, we set the \texttt{Depth} parameter so that a single level of nesting is allowed. Although our model can handle an arbitrary number of nested frames, a single level suffices to discover the minimum-size trace for all queries.

The effects of this optimization are shown in \autoref{tab:evaluation}: we can see that disabling framing for the queries that do not require multilevel DOM trees considerably lowers the solving time.

\bfparagraph{Fixed size arrays.}
Our model uses functional arrays~\cite{McCarthy62,Reynolds79} 
for the definition of the HTML and DOM objects and the implementation of the window/frame tree.
However, functional arrays are known for significantly increasing the
complexity of queries~\cite{SuzukiJ80,StumpBDL01}.
Therefore, in order to ease the resolution, our compiler provides an
optimization which turns functional arrays into arrays of a fixed size chosen
at compile time.
Since choosing a small size could make a query unsolvable, we run in parallel
several instances of the same query with different sizes and keep the first one that succeeds.
Surprisingly, a size of 5 is enough for all the queries except those for
\emph{Safe policy inheritance} (\#7 and \#8 in \autoref{tab:evaluation}) which require a size of 7.

\section{Related Work}
\label{sec:related}

\bfparagraph{Models of the Browser.}
Bohannon~\cite{BohannonPhd} proposed Featherweight Firefox, a model of a Web browser written in Coq for the verification of security properties concerning JavaScript execution.
The model supports several JavaScript features, such as DOM manipulation, XHR requests, event listeners, and code evaluation via the \texttt{eval} function. However, the set of modeled Web components comprises only windows, cookies, and selected HTML tags (\texttt{<script>}, \texttt{<div>}).
Bugliesi \etal{}~\cite{BugliesiCFK15} extended Featherweight Firefox to formalize the security guarantees of \texttt{HttpOnly} and \texttt{Secure} cookies against network and Web attackers able to exploit XSS vulnerabilities.
In~\cite{BugliesiCFKT14} the authors use Featherweight Firefox as a starting point to develop a pen-and-paper model of a security-enhanced browser which enforces a Web session integrity property that captures attacks like CSRF and credential theft via XSS.

In contrast to \framework, Bohannon's model and later extensions were developed with machine-checked proofs in mind and have not been used to automatically detect vulnerabilities. They also lack support for most of the Web features considered in our invariants, e.g., CORS, CSP, service workers.
Because of their focus on Web script security, they formalize script executions using a small-step semantics. This choice allows for a precise modeling of JavaScript but  hinders automatic verification, as it forces solvers to handle JavaScript programs.

\bfparagraph{Models of the Web.}
In their seminal work, Akhawe \etal~\cite{AkhaweBLMS10} developed the first formal model of the Web ecosystem.
The authors encoded in the model a set of security goals, which include fundamental properties of the Web platform that are assumed to hold, and a notion of session integrity capturing CSRF attacks. The validity of these goals has been checked with the Alloy Analyzer~\cite{Jackson02} and their violations pointed out the existence of novel and previously known attacks.

Despite being similar in spirit to our proposal, there are important differences between \framework\ and the model of Akhawe \etal{}
First, the model cannot be used to prove security properties, since the Alloy Analyzer uses SAT-based bounded model checking, but just to disprove them, while \framework\ can be used also to produce automated or machine-checked proofs.
Second, being developed in 2010, it lacks many features of the modern Web (e.g., frames, CSP and service workers) that are a fundamental part of our model. Adding these features a posteriori would not be possible without rewriting the model from scratch, since some of them (\eg, a faithful handling of frames) are core components of Web browsers.
Last, contrary to \framework{}, the model of Akhawe \etal\ is stateless. For this reason,
temporal relations between events, e.g., the correct sequencing of requests and response pairs, need to be explicitly modeled. Considering that Web Standards are typically written using a stateful imperative style, a more natural modeling follows a stateful approach, as employed in our model.

Bansal \etal~\cite{BansalBDM14} developed WebSpi, a generic library that defines the basic components of the Web infrastructure (browsers, HTTP servers) and enables developers to automatically verify security properties of specific Web applications / protocols using ProVerif~\cite{Blanchet01}.

The browser model of WebSpi is rather primitive and includes a subset of the features supported in \framework. This is in line with the intended usage of WebSpi, i.e., the verification of Web protocols, for which it suffices to model only the features used by the protocol under analysis. Instead, we target inconsistencies between Web features themselves, without focusing on a specific Web protocol or application, for which we need a much more comprehensive browser model. Similarly to WebSpi, \framework\ supports automated security proofs: if this does not succeed, however, we can still fall back to machine-checked proofs in Coq.

The most comprehensive and maintained model of the Web to date is the \emph{Web Infrastructure Model} (WIM), a pen-and-paper model which has been used to assess the security of Web Payment APIs~\cite{DoHKSWW22}, Web protocols, e.g., OAuth 2.0~\cite{FettKS16}, OpenID Connect~\cite{FettKS17}, and the Financial-Grade APIs~\cite{FettHK19}.

The browser model of \framework\ supports most of the features of WIM browsers, except for
\begin{enumerate*}[label=(\roman*)]
  \item HSTS, since in our model we abstract away the network,
  \item HTTP basic authentication, because it is an application-specific server-side mechanism,
  \item the Web Payment APIs, since a sensitive usage of this API would require a detailed modeling of the server-side behavior of payment providers and merchants servers.
\end{enumerate*}
On the other hand, \framework\ supports several client-side mechanisms and security policies, like domain relaxation, CSP and CORS, that are not part of WIM.
Additionally, being a pen-and-paper model, WIM can neither be used to automatically discover security vulnerabilities, nor to develop automated or computer-assisted proofs, which are central features of our framework.

\bfparagraph{Other works.}
Quark~\cite{JangTL12} is a WebKit-based Web browser, whose kernel has been implemented and formally verified in Coq. The kernel is responsible for handling input/output
and offers services to the other possibly compromised components of the browser, which deal with various operations such as the rendering of Web pages, handling of cookies and tab management.
The separation of duties between the kernel and browser's components, together with a set of security policies implemented in the kernel, enables Quark developers to formally prove the enforcement  of security properties, such as tab non-interference,  and integrity of cookies and responses. This is orthogonal to our work, which instead aims at devising a formal browser model to validate Web invariants.

Automated testing is a popular methodology employed in software development processes for bug detection. An application of this methodology in the context of Web security is BrowserAudit~\cite{Hothersall-Thomas15}, a framework composed of over 400 automated tests, which can be used to verify the correctness of the implementation of Web security mechanisms in existing browsers. However, BrowserAudit cannot be used to spot bugs at the specification level, which is the goal of our work.

QuickChick~\cite{Paraskevopoulou15,LampropoulosPP18} is a framework for property-based testing written in Coq. It combines formal methods and testing to formally verify that the code of a test generated from a given property is indeed checking its correctness. QuickChick is orthogonal to our work since it focuses on test case generation: \framework{} relies on testing solely to prove the validity of the counterexamples to our invariants produced by the Z3 solver.

\bfparagraph{Summary.}
To conclude, \framework{} is the first framework that mechanizes  formal proofs  and counterexample-finding for Web invariants. In addition, our browser model is the most comprehensive one when it comes to browser-side security mechanisms, as we further detail in Appendix~\ref{sec:appendix-completeness}.

\section{Conclusion}
\label{sec:conclusion}

In this paper we presented \framework{}, the first formal framework for the security analysis of Web platform components that supports the automated detection of logical flaws and allows for the development of machine-checked security proofs.
We showcased the effectiveness of \framework{} by discovering novel attacks and inconsistencies affecting current Web standards, and automatically validated our findings against major browsers. Additionally, we discussed how \framework{} can be used to carry out machine-checked security proofs for vulnerability fixes.
As a future work, besides expanding the model to cover more Web platform components,
we plan to define additional Web invariants by reviewing newly proposed mechanisms and engaging with the community, including developers and editors of Web standards.

\bibliographystyle{IEEEtranS}
\bibliography{IEEEabrv,biblio}

\appendix 
\appendices

\section{Web Invariants}
\label{sec:appendix1}

\subsection{Interactions with the SOP}
\label{sec:cspsop}

With the \texttt{script-src} CSP directive, developers can specify which scripts can be included in a page and thus access the DOM.
This corresponds to the following property.
\begin{invariant}{I.4}
   The DOM of a page protected by CSP can be read/modified only by the scripts allowed by the policy.
\end{invariant}

\noindent
We encode the invariant in \themodel\ as follows:
\begin{lstlisting}[language=coq]
Definition CSPInvariant (gb: Global) (evs: list Event) (st: State) : Prop :=
  forall pt sc ctx pt_u src origin tctx tt _evs,
    Reachable gb evs st ->
    (* A script sc is present in the page *)
    is_script_in_dom_path gb (st_window st) pt sc ctx ->
    (* The DOM of the toplevel window has been modified by sc *)
    evs = (EvScriptUpdateHTML pt (DOMPath [] pt_u) tctx :: _evs ) ->
    (* The toplevel window is protected by CSP *)
    rp_hd_csp (dc_headers (wd_document (st_window st))) = Some
      {| csp_script_src := Some src; csp_trusted_types := tt |} ->
    (* The script sc is allowed by the CSP *)
    origin_of_url (wd_location (st_window st)) = Some origin ->
    csp_src_match src origin (script_src sc).
\end{lstlisting}
Where the \coqe{csp_src_match} predicate holds when the \coqe{src} source expression matches the URL \coqe{script_src sc} in a page loaded from origin \coqe{origin}.

\begin{figure}[t]
  \centering
  \includegraphics[width=0.45\textwidth]{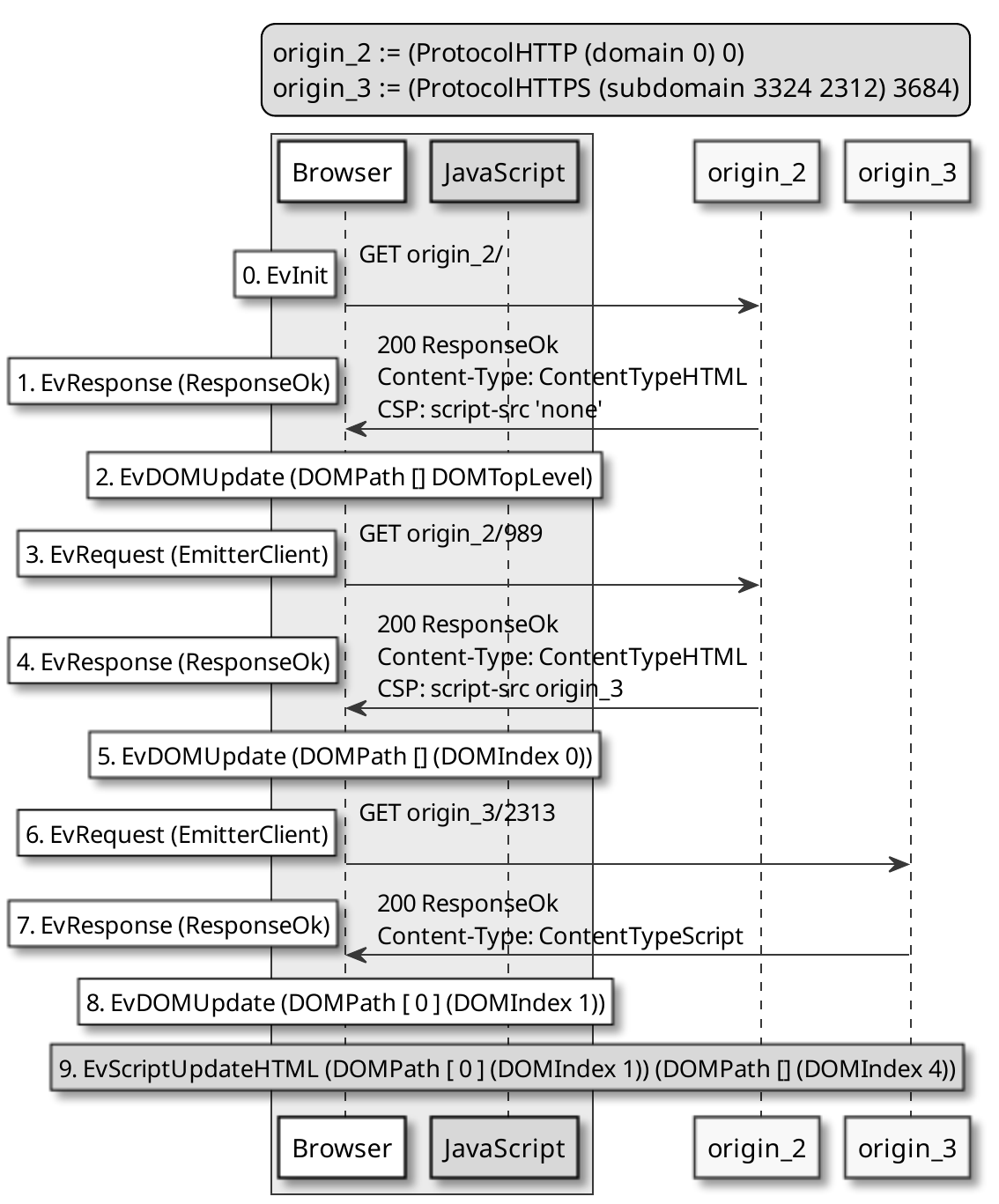}
  \caption{CSP Inconsistency}
  \label{fig:csp}
\end{figure}

\bfparagraph{Attack.}
By running the query, \thetool\ produces a counterexample that corresponds to the CSP violation discovered by the authors of \cite{SomeBR17}.
The complete trace is shown in \autoref{fig:csp}:
a page with \code{Content-Security-Policy: script-src 'none'} (no scripts allowed) contains a same-origin (\code{origin\_2}) iframe with CSP \code{script-src origin\_3}. Then, the iframe loads a script from \code{origin\_3} which alters the DOM of the parent page, violating the Web invariant~\cite{SomeBR17}. Notice that this interaction is allowed by the SOP since both pages share the same origin. 
This violation can be prevented by enforcing the same CSP on all same-origin pages~\cite{OriginPolicyDraft} and by disabling domain relaxation. The Coq proof of the correctness of this solution is available at~\cite{WebSpecSource}.

\subsection{Integrity of server-provided policies}
\label{sec:serviceworkers}

Given the privileged position of service workers, we must ensure that the integrity of security policies attached to server responses is preserved, and such policies are correctly enforced by the browser. This corresponds to the following Web invariant.

\begin{invariant}{I.5}
  If a response from the server contains a security policy, then the browser enforces that specific policy.
\end{invariant}

\noindent We encode the invariant in \themodel\ as follows:
\begin{lstlisting}[language=coq]
Definition SWInvariant (gb: Global) (evs: list Event) (st: State) : Prop :=
  forall corr rq_idx rp_idx rp em,
    Reachable gb evs st ->
    (* Get the server response *)
    is_server_response gb rq_idx rp ->
    (* Get the response that was rendered *)
    in_history (st_fetch_engine st) corr (em,rq_idx,rp_idx) ->
    (* The CSP of the rendered response is equal to the server one *)
    rp_hd_csp (rp_headers rp) =
    rp_hd_csp (rp_headers ((responses gb).[rp_idx])).
\end{lstlisting}
For every response \coqe{rp} that would be generated by the server for a specific request index \coqe{rq_idx},
the response that has been rendered by the browser is present in the \coqe{ft_history} field of the \coqe{FetchEngine}.
In particular, the history stores the mapping between requests and responses (\coqe{rq_idx}, \coqe{rp_idx} at line 7) for every response that is rendered by the browser.
The invariant requires that the CSP of the response that is present in the history must be the same as the one that is generated by the server.

\bfparagraph{Attack.}
We configure the model to avoid a trivial violation of the invariant, i.e., service workers stripping security headers from responses served to the client. This can be prevented by removing the ability of service workers to create synthetic responses via the \code{Request} constructor.

\begin{figure}[t]
  \centering
  \includegraphics[width=0.45\textwidth]{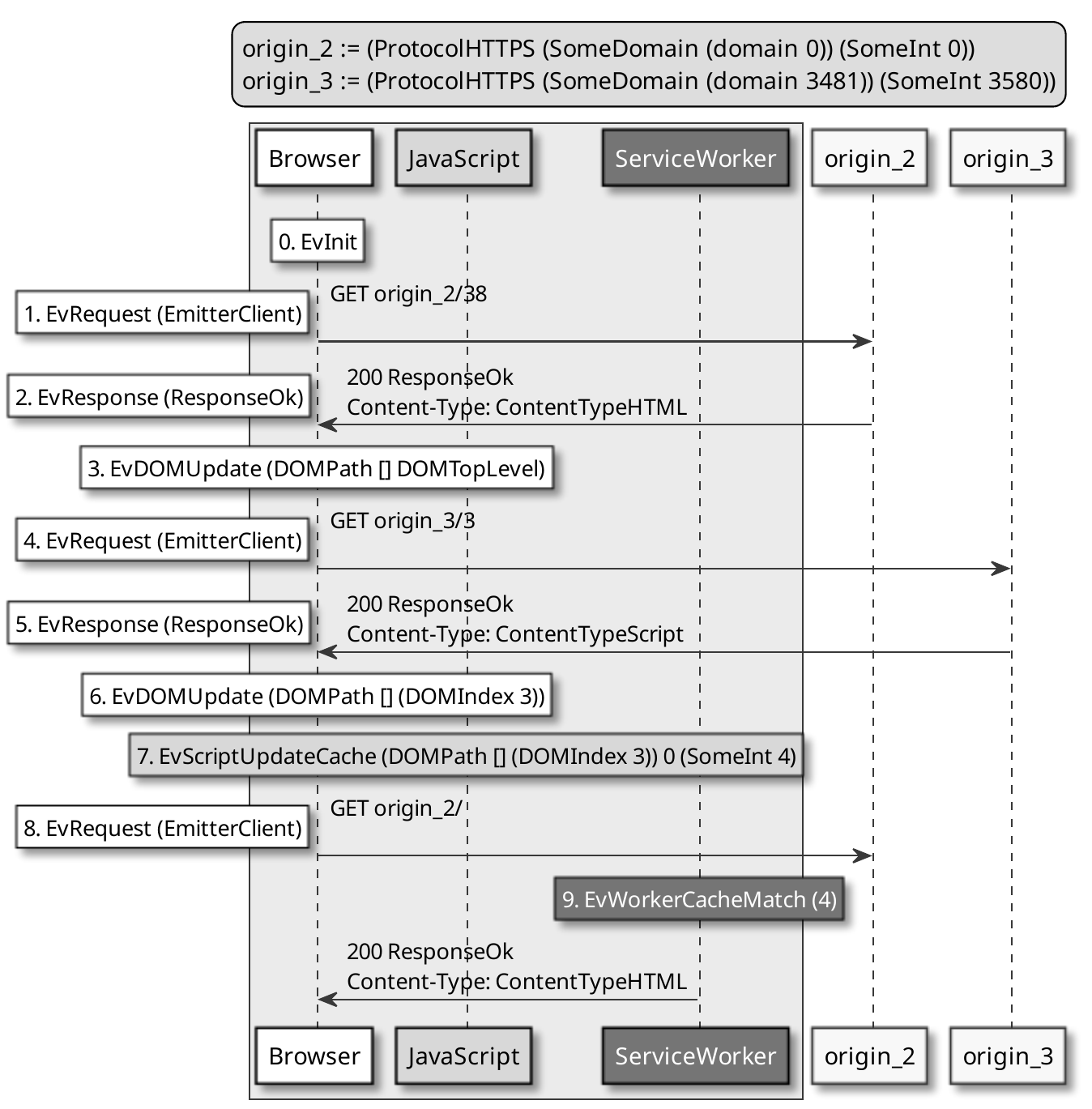}
  \caption{Service Workers Cache Inconsistency}
  \label{fig:swcache}
\end{figure}

Running the query on \theframework{} produces the counterexample shown in \autoref{fig:swcache}:  the invariant is broken when a service worker returns a synthetic response from the cache. Notice that this synthetic response was not generated by the service worker. Instead, it was added to the cache by a script running on the same origin of the service worker (\code{origin\_2}). This flow corresponds to the attack described by Squarcina \etal~\cite{SquarcinaCM21}, where an attacker tampers with cached responses to strip or weaken the CSP served to the user. As the authors pointed out, this issue can be prevented by making the Cache API inaccessible to scripts running in the page context. A security proof of this fix is available online~\cite{WebSpecSource}.

\section{Compiler}
\label{sec:appendix_compiler}

\framework{} includes a compiler that aims to find inhabitants of inductive
types, a  problem which is known to be undecidable for CIC, the logic of
Coq~\cite{DR18}.
To this end, the compiler translates terms in a fragment of CIC into CHC
logic, \ie, first-order logic with fixed-points expressed in terms of
Constrained Horn Clauses, hence discharging the undecidability of the problem
to CHC solvers~\cite{HBM11,HB12}.
In the following, we give an overview of how our compiler performs this
translation.

\subsection{Considered CIC Fragment}

Contrary to related work \cite{Czajka2018, Czajka2020}, our compiler does not
perform a shallow embedding into \emph{untyped} first-order logic, but instead
performs a type-preserving translation into CHC logic, \ie, \emph{typed}
first-order logic with fixed-point.
If, on the one hand, this allows us to leverage all the power of CHC
solvers, this comes, on the other hand, at the price of restrictions on the
fragment of the logic of Coq we consider.

The considered fragment of the logic of Coq we consider is CIC without
dependent types (1), and where inductive type annotations and constructor
arguments are restricted to ground variables (2).
We also require inductive type parameters to be instantiated when the compiler
is called.
Before discussing these limitations, note that the resulting logic is still
extremely expressive as it contains System $F_\omega$, the higher-order
polymorphic lambda calculus.
This also means that the inhabitation problem is still undecidable on this
fragment \cite{DR18}.

The reason for the restriction on inductive type annotations and constructor
arguments (2) is twofold.
The first reason is that CHC solvers do not performs type equation
resolution, and therefore introducing symbolic type variables is
forbidden.
It is possible to circumvent this issue by performing a shallow embedding of
types, however this would likely come  at a significant cost in resolution time.
The second reason is similar to the first, but for functions.
However in this case, we expect this restriction to be relaxed in the future
thanks to recent progress in function synthesis \cite{BarbosaRLT19,
ReynoldsBNBT19}.

The restriction on dependent types (1) could also be circumvented by shallow
embedding, but again at a high cost in resolution time.
Instead, upcoming development of \framework{} aims to relax this restriction
so that dependent types are allowed in inductive types.
This relaxation will cover a significant number of practical cases, like the
famous example where the type of an array includes a program expression giving
the size of that array.

\subsection{Compilation Pipeline}

In order to translate the support fragment of CIC to CHC logic,
our compiler performs the following steps:
\begin{description}[leftmargin=0pt]

  \item[Term-Type-Kind Hierarchy]
    From a syntactic point of view, types cannot be distinguished from terms in
    CIC.
    Because CHC does not permit such intricacy, we have to built a strict
    term-type-kind stratified hierarchy \cite{Barendregt92}, where kinds are
    defined as
    $k := \mbox{Prop}\ |\ \mbox{Set}\ |\ k \rightarrow k$.
    This stratification is done by recursive exploration, starting from the
    inductive type on which the compiler is called, and following CIC typing
    rules\footnote{\url{https://coq.inria.fr/refman/language/cic.html}} to
    deduce to which stratum each syntactic term belongs.
    We rely on Coq type-checking to ensure that connections between terms,
    types and kinds are sound.
    As a side effect, this stratification makes a clear distinction between
    types and proposition or between terms and proofs, which will ease
    subsequent steps.

  \item[Partial Application]
    In CIC, any term can be partially applied.
    This includes functions of course, but also inductive types, constructors, or
    type definitions.
    Such flexibility is not allowed in CHC, and therefore all partial
    applications have to be removed.
    This is done by systematically performing $\eta$-expansion~\cite{DanvyP96}
    on every term that could be applied.

  \item[Lambda Abstraction]
    We also have to removed lambda abstractions, both those which are present
    in the original CIC terms and those which were introduced by
    $\eta$-expansion.
    To this end, we perform $\beta$-reduction wherever possible and
    remove remaining lambda-abstractions by lambda-lifting~\cite{Johnsson85,
    MorazanS07}.

  \item[Polymorphism and Higher-Order]
    Thanks to the previous steps, all functions are now defined at top-level and
    totally applied.
    Therefore we can now remove the use of polymorphism and higher-order simply
    by specialization:
    For every application of a function (resp.\ an inductive type) to a type or
    a function argument, we generate a specialized version of the function
    (resp.\ the inductive type) where the type or function parameter is
    replaced by the argument.

  \item[Constructor Constraints]
    Constructors of inductive types in CIC can contain terms with arbitrary
    constraints, while CHC only supports simple algebraic datatypes.
    Therefore, we split every non-simple inductive type into a simple inductive
    type of kind Set and an inductive type of kind Prop which encapsulates
    these constraints.

\end{description}

Once these steps are  done, the rest of the compilation is straightforward.
Simple inductive types of kind Set are mapped to CHC algebraic datatypes,
inductive types of kind Prop are mapped to relations,
while CIC terms, types and proposition are mapped to CHC terms, sorts, and
formulas.

\section{Scalability}
\label{sec:appendix-scalability}
In this section, we report on the result of the experimental evaluation of the scalability of our browser model. In particular, we measure how the addition of individual Web components affects the performance of \framework\ counterexample finding pipeline, and show how lemmas (\autoref{sec:evaluation-lemmas}) are the most effective tool for improving the solving time. We focus, as our main case study, on the \textit{Integrity of server-provided policies} Web invariant (\autoref{sec:serviceworkers}), since the counterexample found by \framework\ requires the browser model to only support service workers, the cache API, and the CSP header, outside of the core browser features (e.g., requests, responses, DOM, etc). 

Starting from the features listed in \autoref{tab:completeness}, we identify 10 modules, each representing a Web platform feature and refactor our model to be configurable w.r.t. the included components.
With this modification our model is composed of a (core) core set of browser functionality on top of which we are able to automatically include or exclude \begin{itemize*}
    \item[(cookies)] the \code{Cookie} and \code{Set-Cookie} headers, the cookie jar, and the \code{document.cookie} JavaScript API;
    \item[(redir)] HTTP redirections and response codes;
    \item[(cors)] the CORS protocol and its request and response headers;
    \item[(csp)] the \code{Content-Security-Policy} headers and rules, including the \code{script-src} and \code{trusted-types} directives;
    \item[(sw)] service Workers and the JavaScript Cache API;
    \item[(lsc)] support for local scheme URLs, the \code{URL.createObjectURL} API, and the inheritance rules for CSP;
    \item[(ref)] the \code{Referer} request header and the referrer policy mechanism;
    \item[(pmsg)] the Web messaging API (\code{window.postMessage});
    \item[(lst)] the local storage API (\code{window.localStorage}).
\end{itemize*}

Table \ref{tab:scalability} reports the time required by WebSpec to find the counterexample for our case study on 8 incrementally more complex versions of the browser model. For each version, we include one additional features and run the query twice, measuring the running time with or without the inclusion of lemmas. 
The table clearly confirm the intuition that the solving time increases with the addition of new features, as the base model requires less than one hour, compared to the ~16 hours of the complete model. This increase however is not predictable, as the addition of a feature may benefit the solver by introducing constraints which limit the search space, as for example in line \#4, where the addition of CORS improves the solving time by one third.
In all versions of the model, lemmas offer a substantial improvement of the counterexample finding pipeline, never exceeding the 3 minutes of solving time. In our case study, the lemma which is first applied by the solver is \code{script\_state} (\autoref{sec:evaluation-lemmas}), which comprises the first 5 events of the 9 present in the complete trace. This shows that limiting the number of steps the solver needs to consider brings a more noticeable improvement than the simplification obtained by the removal of Web components.

\begin{table}[t]
    \caption{Solving time of invariant \#4 (\autoref{sec:serviceworkers}) for progressively more complex models}\label{tab:scalability}
    \scriptsize
    \centering
    \begin{tabularx}{\linewidth}{X|m{7pt}m{7pt}m{7pt}m{7pt}m{7pt}m{7pt}m{7pt}m{7pt}|rr}
    \toprule
    \multirow{2}{*}{\textbf{\#}} & \multicolumn{8}{c|}{\textbf{Features}} & \multicolumn{2}{c}{\textbf{Solving Time}} \\
    & {\tiny base} & {\tiny cookie} & {\tiny redir} & {\tiny cors} & {\tiny lsc} & {\tiny ref} & {\tiny pmsg} & {\tiny lst} & Baseline & w/ Lemma \\
    \midrule
    1 & \cyes & \cno & \cno & \cno & \cno & \cno & \cno & \cno & 71m & 13s \\
    2 & \cyes & \cyes & \cno & \cno & \cno & \cno & \cno & \cno & 3h 34m & 58s \\
    3 & \cyes & \cyes & \cyes & \cno & \cno & \cno & \cno & \cno & 3h 25m & 1m 7s \\
    4 & \cyes & \cyes & \cyes & \cyes & \cno & \cno & \cno & \cno & 2h 14m & 48s \\
    5 & \cyes & \cyes & \cyes & \cyes & \cyes & \cno & \cno & \cno & 6h 54m & 2m 11s \\
    6 & \cyes & \cyes & \cyes & \cyes & \cyes & \cyes & \cno & \cno & 9h 32m & 1m 52s \\
    7 & \cyes & \cyes & \cyes & \cyes & \cyes & \cyes & \cyes & \cno & 13h 20m & 2m 7s \\
    8 & \cyes & \cyes & \cyes & \cyes & \cyes & \cyes & \cyes & \cyes & 15h 46m & 3m \\
    \bottomrule
    \end{tabularx}
    \ \\ \ \\
{\scriptsize \centering
base: Core Browser Functionality, CSP and Service Workers; 
cookie: Cookies;
redir: HTTP Redirections; cors: CORS Protocol and Headers; 
lsc: Local Schemes; ref: Referer and Referrer Policy;
pmsg: Post Message; lst: Local Storage }
\end{table}

\section{Completeness}
\label{sec:appendix-completeness}

\autoref{tab:completeness} provides an overview of the features supported by the models discussed in \autoref{sec:related}. In particular, here we focus exclusively on Web components implemented in Web browsers, since this is ultimately the goal of our work, but models like WIM~\cite{DoHKSWW22}, WebSpi~\cite{BansalBDM14} and Alloy~\cite{AkhaweBLMS10} additionally implement a variety of other features that are needed to model other parts of the Web ecosystem.

One of the main differences among the different proposals lies in the way JavaScript is modeled. WIM and Featherweight Firefox model the small-step semantics of (a subset of) JavaScript: this is of fundamental importance, \eg, in WIM, since the model has been used to verify the security of Web protocols and it is necessary to define the precise semantics of script used by the parties involved in the protocol run.
In \framework, similarly to the Alloy model of Akhawe \etal~\cite{AkhaweBLMS10}, we are only interested to the API calls that a script can perform. For this reason, we do not specify the exact behavior of a script, rather we assume that a script can call the supported APIs in any arbitrary way, using any data in its knowledge as parameters to these calls.

From the perspective of features support, \framework{} and WIM are the two most complete models available so far. As mentioned in \autoref{sec:related}, \framework{} does not support HSTS, HTTP basic authentication and the Web Payment API, since we abstract away from the network and from the specific implementation of Web servers.
On the other hand, we support a variety of features that are missing in WIM browsers, such as CORS, cookie attributes like \code{Domain}, \code{Path} and \code{SameSite}, the \code{\_\_Host-} prefix, CSP, the Cache API, interception of request of service workers, the \code{document.cookie} API, which play a prominent role for many of the attacks reported in this paper.
As shown in the table, there are also some minor differences concerning modeled URL components, type of supported HTTP redirects, status codes and headers, and functionality of service workers.

Concerning the other models, \framework{} essentially supports all the features implemented by them. Currently we only support the \code{Origin} and \code{Access-Control-Allow-Origin} HTTP headers for CORS, while the Alloy model supports all of them\footnote{For space reasons, in \autoref{tab:completeness} we let \code{ACA} stand for \code{Access-Control-Allow}, \code{AC} for \code{Access-Control} and \code{ACR} for \code{Access-Control-Request}.}: although the headers supported in \framework{} are sufficient to implement the fundamental CORS functionalities, the remaining ones allow a more careful treatment of CORS and we plan to implement them as future work.

\newcolumntype{C}{>{\hsize=0.2\hsize\raggedleft\arraybackslash}X}%

\renewcommand{\code}[1]{\texttt{\hyphenchar\font=`\-#1}}

\begin{table*}
	\caption{Comparison of supported Web components in existing models}\label{tab:completeness}	
	\tiny
	\begin{tabularx}{\linewidth}{p{1.5cm}p{4cm}cccccc}
	    \toprule
	    \multicolumn{2}{c}{\textbf{Web Components}} & \textbf{WebSpec} & \textbf{WIM}~\cite{DoHKSWW22} & \textbf{WebSpi}~\cite{BansalBDM14} & \textbf{Alloy}~\cite{AkhaweBLMS10} & \textbf{FF}~\cite{BohannonPhd} \\
	    \midrule
	    
	    \multirow{9}{*}{\textbf{URLs}} & Scheme & \cyes & \cyes & \cyes & \cyes & \cyes \\
	    & \quad HTTP(S) & \cyes & \cyes & \cyes & \cyes & \cyes \\ 
	    & \quad Pseudo-protocols & \code{data}:, \code{blob}: & - & - & - & \code{about}: \\
	    & Host & \cyes & \cyes & \cyes & \cyes & \cyes \\
	    & Port & \cyes & \cno & \cno & \cno & \cno \\
	    & Path & \cyes & \cyes & \cyes & \cyes & \cyes \\
	    & Parameters & \cno & \cyes & \cyes & \cyes & \cyes \\
	    & Fragment & \cno & \cyes & \cno & \cno & \cno \\
	    & JS API \code{URL.createObjectURL} & \cyes & \cno & \cno & \cno & \cno \\
	    \midrule
	    
        \multirow{18}{*}{\textbf{HTTP}} & Request methods &&&&& \\
        & \quad \code{GET} & \cyes & \cyes & \cyes & \cyes & \cyes \\
        & \quad \code{POST} & \cyes & \cyes & \cyes & \cyes & \cno \\
        & \quad Others & \code{PUT}, \code{DELETE}, \code{OPTIONS} & \parbox{2cm}{\centering \code{PUT}, \code{DELETE}, \code{OPTIONS}, \code{TRACE}, \code{CONNECT}} & - & \code{PUT}, \code{DELETE}, \code{OPTIONS} & - \\
        & Response codes &&&&& \\
        & \quad Redirection & 302, 307 & 303, 307 & 302 & 302, 303, 307 & - \\
        & \quad Others & 200, 204 & 101, 200 & 200 & 200, 401 & 200 \\
        & Headers (not fitting the categories below) &&&&& \\
        & \quad \code{Referer} & \cyes & \cyes & \cno & \cno & \cno \\
        & \quad \code{ReferrerPolicy} & \cyes & \cyes & \cno & \cno & \cno \\
        & \quad \quad Directive \code{origin} & \cyes & \cyes & \cno & \cno & \cno \\
        & \quad \quad Directive \code{no-referrer} & \cyes & \cyes & \cno & \cno & \cno \\
        & \quad \quad Directive \code{unsafe-url} & \cyes & \cyes & \cno & \cno & \cno \\
        & \quad \code{Authorization} & \cno & \cyes & \cno & \cno & \cno \\
        & \quad \code{Content-Type} & \cyes & \cno & \cno & \cno & \cno \\ 
        & \quad \code{Location} & \cyes & \cyes & \cyes & \cyes & \cno \\
        & \quad \code{Strict-Transport-Security} & \cno & \cyes & \cno & \cno & \cno \\
        & \quad \code{WWW-Authenticate} & \cno & \cno & \cno & \cyes & \cno \\
	    \midrule
	    
        \multirow{13}{*}{\textbf{Cookies}} & HTTP headers &&&&& \\
	    & \quad \code{Cookie} & \cyes & \cyes & \cyes & \cyes & \cyes \\
	    & \quad \code{Set-Cookie} & \cyes & \cyes & \cyes & \cyes & \cyes \\
	    & Attributes &&&&& \\
	    & \quad \code{Domain} & \cyes & \cno & \cno & \cyes & \cyes \\ 
	    & \quad \code{Path} & \cyes & \cno & \cno & \cyes & \cyes \\ 
	    & \quad \code{Secure} & \cyes & \cyes & \cno & \cyes & \cno  \\ 
	    & \quad \code{HttpOnly} & \cyes & \cyes & \cno & \cno & \cno  \\ 
	    & \quad \code{SameSite} & \cyes & \cno & \cno & \cno & \cno \\
	    & \code{\_\_Secure-} Prefix & \cyes & \cyes & \cno & \cno & \cno \\
	    & \code{\_\_Host-} Prefix & \cyes & \cno & \cno & \cno & \cno \\
	    & JS API \code{document.cookie} & \cyes & \cyes & \cyes & \cno & \cyes \\ 
	    & SOP for cookies & \cyes & \cyes & \cyes & \cyes & \cyes \\
	    \midrule
	    
        \multirow{6}{*}{\textbf{Windows}} & Multiple tabs & \cno & \cyes & \cno & \cno & \cyes \\
	    & Framing support & \cyes & \cyes & \cno & \cno & \cno \\
	    & Cross-window communication (postMessage API) & \cyes & \cyes & \cno & \cno & \cno \\
	    & \quad JS API \code{window.location} & \cyes & \cyes & \cno & \cno & \cno \\
	    & \quad JS API \code{window.history} & \cno & \cyes & \cno & \cno & \cno \\
	    & \quad JS API \code{window.close} & \cno & \cyes & \cno & \cno & \cno \\
	    \midrule
	    
        \multirow{4}{*}{\textbf{DOM}} & Supported elements & \parbox{2cm}{\centering \code{<script>}, \code{<iframe>}, \code{<form>}, \code{<img>}} & \parbox{2cm}{\centering \code{<script>}, \code{<iframe>}, \code{<form>}} & - & \code{<form>} & \code{<script>}, \code{<div>} \\
        & JS API for DOM manipulation & \cyes & \cyes & \cno & \cno & \cyes \\ 
        & JS API \code{document.domain} & \cyes & \cno & \cno & \cno & \cno \\
        & SOP for DOM access & \cyes & \cyes & \cno & \cno & \cyes \\
        \midrule
	    
        \multirow{4}{*}{\parbox{1.5cm}{\textbf{XMLHttpRequest / Fetch API}}} & SOP for XHR / fetch requests & \cyes & \cyes & \cyes & \cyes & \cyes \\
        & Sending requests via JavaScript & \cyes & \cyes & \cyes & \cyes & \cyes \\
        & Reading responses via JavaScript & \cyes & \cyes & \cyes & \cno & \cyes \\
        & \quad Forbidden response headers (\code{Set-Cookie}) & \cyes & \cyes & \cno & \cno & \cno \\
	    \midrule
	    
        \multirow{8}{*}{\textbf{CORS}} & Request types &&&&& \\
        & \quad Simple requests & \cyes & \cno & \cno & \cyes & \cno \\
        & \quad Non-simple requests (w. preflight) & \cyes & \cno & \cno & \cyes & \cno \\
        & HTTP headers &&&&& \\
        & \quad \code{Origin} & \cyes & \cyes & \cyes & \cyes & \cno \\
        & \quad \code{Access-Control-Allow-Origin} & \cyes & \cno & \cno & \cyes & \cno \\
        & \quad Others & - & - & - & \parbox{2.5cm}{\centering \code{ACA-Method}, \code{ACA-Headers}, \code{ACA-Credentials}, \code{AC-Max-Age}, \code{ACR-Method}, \code{ACR-Headers}} & - \\
	    \midrule

        \multirow{8}{*}{\parbox{1.5cm}{\textbf{CSP /\\Trusted Types}}} & CSP directives &&&&& \\
        & \quad \code{script-src} & \cyes & \cno & \cno & \cno & \cno \\
        & \quad \code{trusted-types} & \cyes & \cno & \cno & \cno & \cno \\
        & \quad \code{require-trusted-types-for} & \cyes & \cno & \cno & \cno & \cno \\
        & CSP Inheritance & \cyes & \cno & \cno & \cno & \cno \\
        & Trusted types &&&&& \\
        & \quad Create trusted types (\code{policy.createHTML}) & \cyes & \cno & \cno & \cno & \cno \\
        & \quad Secure context restriction & \cyes & \cno & \cno & \cno & \cno \\
	    \midrule
	    
        \multirow{4}{*}{\textbf{Service Workers}} & Interception of requests (\code{evt.respondWith}) & \cyes & \cno & \cno & \cno & \cno \\
        & Access to the cache API & \cyes & \cno & \cno & \cno & \cno \\
        & Messaging with other windows & \cno & \cyes & \cno & \cno & \cno \\
        & Opening new windows (\code{Clients.windowOpen}) & \cno & \cyes & \cno & \cno & \cno \\
	    \midrule
	    
        \multirow{7}{*}{\textbf{Storage APIs}} & Cache API \\
	    & \quad \code{Caches.put} & \cyes & \cno & \cno & \cno & \cno \\
	    & \quad \code{Caches.match} & \cyes & \cno & \cno & \cno & \cno \\
	    & \quad Secure context restriction & \cyes & \cno & \cno & \cno & \cno \\
	    & Local storage \\
	    & \quad \code{localStorage.getItem} & \cyes & \cyes & \cno & \cno & \cno \\
	    & \quad \code{localStorage.setItem} & \cyes & \cyes & \cno & \cno & \cno \\
	    \midrule
	    
	    \textbf{Web Payment APIs} & & \cno & \cyes & \cno & \cno & \cno \\
		\bottomrule
	\end{tabularx}
\end{table*}

\end{document}